\documentclass[twocolumn,showpacs,prb,superscriptaddress,aps,floatfix]{revtex4-1}
\usepackage{rotating}
\usepackage{amsmath}
\usepackage{color}
\usepackage{graphicx}
\usepackage{graphics}
\usepackage{epsfig}
\usepackage{comment}
\usepackage{gensymb}
\def\ai         {{\it ab-initio}\,\,}
\def\ga         {\alpha}

\def\gD         {\Delta} 
\def\gee        {\varepsilon}
\def\gl         {\lambda}
\def\gL         {\Lambda}
\def\go         {\omega}
\def\goql       {\omega_{\qq \gl}}
\def\gql        {\qq \gl}

\def\la         {\langle}
\def\ra         {\rangle}
\def\kk         {{\bf k}}
\def\qq         {{\bf q}}

\def\rr	     	{{\bf r}}

\def\Vscf       {V_\text{scf} }

\def\ep         {electron-phonon }

\def\gsq        {{\mid g^{\gql}_{n' n \kk} \mid}^2}

\def\nk         {n\kk}

\def\npk        {n' \kk}
\def\rnk        {\mid  n     \kk    \ra}

\def\uot        {\frac{1}{2}}

\usepackage{float}

\begin{document}
\title{Thermal evolution of silicon carbide electronic bands}
\author{E. Cannuccia}
\affiliation{Aix-Marseille Universit\'e , Laboratoire de Physique des Interactions Ioniques et Moléculaires (PIIM), UMR CNRS 7345, F-13397 Marseille, France}
\affiliation{Dipartimento di Fisica, Universit\`a di Roma ``Tor Vergata'', Via della Ricerca Scientifica 1, I--00133 Roma, Italy}
\author{A. Gali}
\affiliation{Wigner Research Centre for Physics, PO.\ Box 49, H-1525, Budapest, Hungary}
\affiliation{Department of Atomic Physics, Budapest University of Technology and Economics, Budafoki \'ut 8., H-1111 Budapest, Hungary}

\begin{abstract}
Direct observation of temperature dependence of individual bands of semiconductors for a wide temperature region is not straightforward, in particular. However, this fundamental property is a prerequisite in understanding the electron-phonon coupling of semiconductors. Here we apply \emph{ab initio} many body perturbation theory to the electron-phonon coupling on hexagonal silicon carbide (SiC) crystals and determine the temperature dependence of the bands. We find a significant electron-phonon renormalization of the band gap at 0~K. Both the conduction and valence bands shift at elevated temperatures exhibiting a different behavior. We compare our theoretical results with the observed thermal evolution of SiC band edges, and discuss our findings in the light of high temperature SiC electronics and defect qubits operation. 
\end{abstract}
 \maketitle
\section{Introduction}
Electron-phonon interaction impacts a large variety of fundamental materials properties~\cite{RevModPhys.89.015003}, from the critical temperature of superconductors to the zero-point renormalization and the temperature dependence of the electronic energy bands, from the electronic band gaps~\cite{Cardona1989,Cardona2005,Cannuccia2011,Cannuccia_epjb,Gali2016} to the thermal evolution of the optical spectra and excitonic lifetimes~\cite{Marini2008, Kawai2014, Sanchez2016}. In addition the electron-phonon coupling contributes to the optical absorption and emission in indirect gap semiconductors~\cite{Noffsinger2012,Paleari2019,Cannuccia2019}, determines the electronic carrier mobility of semiconductors~\cite{Ponce2018}, the carrier relaxation rates~\cite{MolinaSanchez2017}, the distortion of band structures and phonon dispersion giving rise to kinks and Kohn anomalies in photoemission~\cite{Forster2013}.

The thermal evolution of the band structure and band gap arises from the thermal expansion effect and from the coupling of electrons with phonons leading to a renormalization of the electronic states. The latter effect is in most of the cases the dominant one, being larger than the thermal expansion~\cite{Cardona1989}. The strength of renormalization depends on temperature so that valence and conduction bands may be shifted differently leading usually to a shrinking of the gap~\cite{Cardona1989}, although an anomalous behaviour (i.e. gap increases with temperature) is found in other cases~\cite{Villegas2016}.

\begin{figure}
\includegraphics[width=0.4\textwidth]{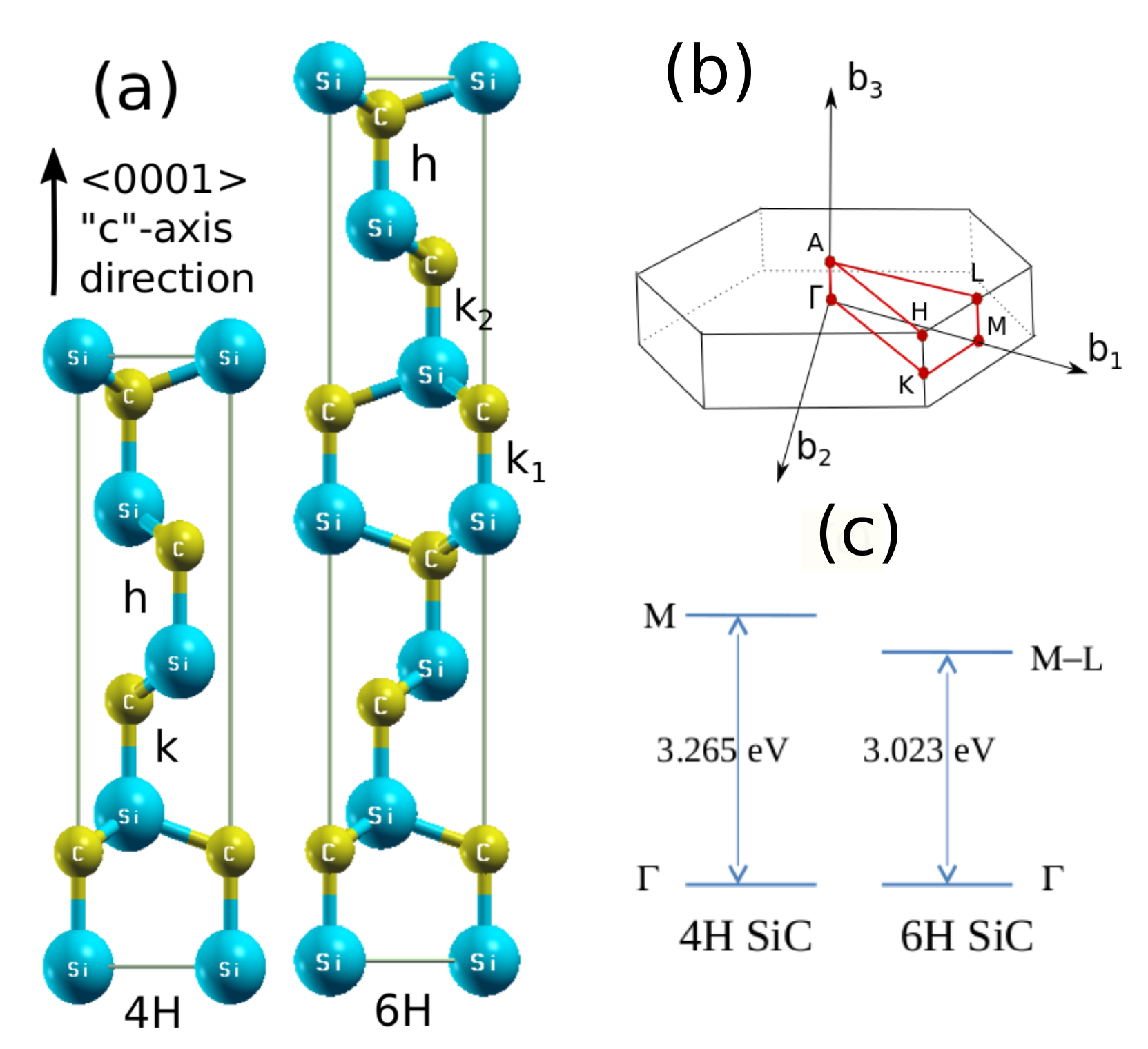}
\caption{\label{fig:SiC}(a) Primitive cells of $4H$ and $6H$ SiC with $k$ ($k_1$, $k_2$) and $h$ Si-C bilayers, where $k$ and $h$ refers to quasicubic and hexagonal sites. (b) Brillouin-zone of hexagonal SiC polytypes. (c) Sketch of band gap of hexagonal SiC crystals. In $4H$ SiC, the first ($M_1$) and second ($M_2$) conduction bands are close in energy~\cite{Klahold2017}. In $6H$ SiC, the lowest conduction band is very flat along the $M-L$ line~\cite{Kackell1994, Lambrecht2001}, and it should be close to the $M$ point [c].} 
\end{figure}

Direct observation of the temperature evolution of individual bands over a wide region of temperatures is not straightforward. Optical techniques are capable of measuring band gaps, and not the absolute values of the valence band maximum (VBM) and the conduction band minimum (CBM) separately. The interpretation of results from optical techniques is then weakly conclusive. In addition, the indirect band gap nature of some materials prohibits the direct optical transition between VBM and CBM, which turns to be allowed only when phonons assist the optical excitation.  Recent attempts used Si $2p$ core level as a reference to extract the CBM and VBM energy position of Si and hexagonal $6H$ silicon carbide (SiC) crystals [see Fig.~\ref{fig:SiC}(a)] from the onset of soft X-ray absorption spectroscopy (XAS) and soft non-resonant X-ray emission spectroscopy (XES), respectively~\cite{Beye2010,Miedema2014}. This method assumes temperature independent core exciton binding energy~\cite{Buczko2000}, which results in a systematically smaller derived band gap than the observed optical band gap, and it may suffer from the accurate observation of the onset energies at elevated temperatures caused by temperature broadening effects. We stress that none of these methods enables the observation of individual bands other than band edges but observation of the temperature dependence of those bands can be an important issue at high temperatures.  
It is utterly important to apply \emph{ab initio} many body perturbation theory that can provide valuable insights on the electron-phonon coupling effect in semiconductors by directly describing the temperature dependence of individual bands. This fundamental property has been recently studied
typically, only for the band edges and up to room temperature~\cite{Marini2008,Cannuccia2011, Cannuccia_epjb}. 
Here we are interested to extend these investigations to higher temperatures for hexagonal $4H$ and $6H$ SiC crystals [see Fig.~\ref{fig:SiC}(a)] and in particular we aim to get the temperature dependence of the first and second conduction band in order to establish the effect of the latter on the conductivity.
The choice of these materials was motivated by three points: (i) experimental data are available for the temperature dependence of $4H$ SiC gap~\cite{Choyke1964} as well as $4H$ SiC based semiconductor devices have been successfully tested at high temperatures (around 800~K) for Venus mission~\cite{Neudeck2016}, where deep insight into the electron-phonon interaction of electronic bands has an uttermost importance, (ii) since both experimental data for temperature dependence gap~\cite{Choyke1969} and band edges~\cite{Miedema2014} of $6H$ SiC are available, this compound is eligible for validation of the theoretical methods which rely on the electron-phonon renormalization of electronic states, and (iii) SiC is a semiconductor platform for hosting hybrid opto-electro-mechanical defect quantum bits~\cite{Weber10, Gali:PSSB2011, Koehl11, Falk2014, Udvarhelyi2018, Whiteley2019,  Riedel2012, Simin2016, Cochrane2016, Anisimov2016, Niethammer2019}. These defect quantum bits require very accurate electrical and optical control which depend on the ionization thresholds, i.e., the position of band edges at the operation temperature~\cite{Wolfowicz2017, Golter2017, Magnusson2018}.

\section{Computational Methods}
We perform geometry optimization, electronic structure and phonons calculation, followed by the calculation of electron-phonon matrix elements, and renormalization of the electronic energies as a function of temperature [see Supplemental Material (SM) Ref.~\onlinecite{SM} for further details].
Density functional theory (DFT) and density functional perturbation theory (DFPT) calculations are carried out using {\tt Quantum ESPRESSO 6.1.0}~\cite{QE-2017} suite. We employ norm-conserving pseudopotentials and the exchange-correlation functional is described by the local density approximation (LDA) with the Perdew-Zunger parametrization~\cite{Perdew81}. We perform in a first step a full optimization (starting from experimental lattice parameters and atomic positions along $\vec c$ axis~\cite{Stockmeier2009,Kackell1994}) using 90~Ry as kinetic energy cutoff, $18\times18\times6$ and $18\times18\times4$ k-meshes for $4H$ and $6H$ SiC crystals, respectively. 
The phonon frequencies are obtained with the same kinetic energy cutoff on a $10\times10\times3$ and $10\times10\times2$ q-meshes, respectively, and then interpolated along the q-path connecting $\Gamma-K-M-\Gamma$ high symmetry points of hexagonal systems Brillouin zone [see Fig.~\ref{fig:SiC} (b) and SM for symmetry analysis of phonon modes, phonon dispersion curves and comparison between theoretical and experimental phonon frequencies at $\Gamma$ point for $4H$ and $6H$ SiC~\cite{Patrick1968,Feldman1968,Feldman1968bis,Nienhaus1995,Nakashima1997,Bluet1999}]. 

The same k-point sampling of the Brillouin-zone and cutoff are used to calculate the derivatives of the self-consistent Kohn-Sham potential with respect to the atomic positions needed to evaluate the electron-phonon coupling matrix elements. A denser q-grid ($12\times12\times3$ and $12\times12\times3$ q-meshes for $4H$ and $6H$ SiC, respectively) results in a difference of less than $10\%$, thus we estimate it as an upper bound of our accuracy. Unoccupied bands as many as five times the number of the occupied ones are taken into account for $4H$ SiC. On the other hand for $6H$ we are bound to use 96 unoccupied bands (four times the number of occupied ones) because of technicalities in the applied algorithms and computational capacity. We note that this affects the convergence of $6H$ SiC results as discussed in SM. We will present the extrapolated convergent values for $6H$ SiC that are estimated from the convergence study of $4H$ SiC results. 

\section{Theoretical Background}
The key issue of this study is the calculation of the temperature dependent correction to the electronic state $\rnk$, with energy $\gee_{n\kk}$ due to the electron-phonon interaction. The electron-phonon interaction is treated perturbatively~\cite{ponce2014,marini2015} within Heine Allen Cardona (HAC) approach~\cite{allen_heine1976,allen_cardona_1981,allen_cardona_1983} as implemented in {\tt Yambo}~\cite{Sangalli_2019} code, by considering the first- and second-order Taylor expansion of the self-consistent potential $\Vscf(\rr)$~\cite{DFPT} in the nuclear displacements ${\bf u}_{Is}$ with respect to the equilibrium positions ${\bf R}_{Is}={\bf R}_I + {\bf \tau}_s$ for the atom $s$ inside the cell I (the cell is located at position $R_{I}$) at the position ${\bf \tau}_s$. Standard perturbation theory is then applied. The first order Taylor expansion of $\Vscf(\rr)$ is treated within the second--order perturbation theory, while the second order Taylor expansion is treated within the first-order perturbation theory. The corresponding temperature dependent energy shift of the electronic state is then composed of Fan and Debye-Waller (DW) contributions.    



\begin{table*}[t]
\caption{\label{tab1:energies}Energies of the lowest conduction band (first line) and the highest valence band (second line) at high symmetry points in the hexagonal BZ with different levels of theory compared to the experimental data. The energy bands are referred to the valence band maximum at $\Gamma$ point, which is set to zero. For $M$ point we give the first and second lowest conduction band positions. In addition, the indirect band gap ($E_g^\text{ind}$) is reported together with the optical gaps. The quasiparticle-level correction for the direct gap at $\Gamma$ is also given in the last column.}
\begin{ruledtabular}
\begin{tabular}{cccccccc}
Polytype   & $\Gamma$  &  M                & DFT-LDA $E_g^\text{ind}$  & \multicolumn{2}{c}{GW $E_g^\text{ind}$} & Optical gap\footnote{Ref.~\onlinecite{Landolt-Bornstein,Choyke1969}} &  GW corr. (dir)  \\
           & (eV)      &  (eV)             &   (eV)           & this work (eV)                   &   other works\footnote{Ref.~\onlinecite{Ummels1998}} (eV)                              &     (eV) &      (eV)      \\   \hline
4H         &  5.18    &  2.25 \,\,  2.34  &   2.25           & 3.17                   &     3.35                      &  3.27    &       0.97    \\
           &  0.00    & -1.14             &                  &      &                                     &          &                     \\  
6H         &  5.27    &  2.04 \,\,  2.21  &   2.04           & 2.96                   &    3.24                       &  3.02    &       0.98    \\
           &  0.00    & -1.11             &                  &     &                                      &          & \\
\end{tabular}
\end{ruledtabular}
\end{table*}

The Fan term is given by 
\begin{align}
\gD \gee^{Fan}_{\nk}(T)=\sum_{\gql} \frac{1}{N}\sum_{n'} \frac{\gsq}{\gee_{\nk}-\gee_{n'\kk+\qq}} \times (2 B(\goql) +1),
\label{eq:HAC_Fan_correction}
\end{align}
where $\goql$ are the phonon frequencies calculated \ai using DFPT, while $B(\goql)$ is the Bose function distribution, $\gee_{\nk}$ are the DFT bare electronic energies, $N$ is the number of $\qq$ points, and $\gsq$ are the \ep matrix elements for the scattering between the electronic states $|n\kk \rangle$ and $|n'\kk+\qq \rangle$ via the phonon $\qq\lambda$, defined as  
\begin{multline}
g^{\gql}_{n' n \kk}=\sum_{s \ga} (2 M_s \go_{\gql})^{-1/2} e^{i\qq\cdot\tau_s} \times \\
\times \la n'\kk+\qq | \frac{\partial \Vscf(\rr)}{\partial{R_{s\ga}}}| n \kk \ra \xi_{\ga}(\qq \gl|s),
\label{eq:gkkq}
\end{multline}
where $M_s$ is the atomic mass, $\tau_s$ is the position of the atomic displacement in the unit cell and $\xi_{\ga}(\qq \gl)$ are the Cartesian components $\alpha$ of the phonon polarization vectors corresponding to the phonon momentum $\qq$ and branch $\gl$. We have also used the short form $R_{s\alpha} = R_{Is\alpha}|_{I=0}$.

The DW contribution reads as

\begin{align}
\gD \gee^{DW}_{\nk}(T)=-\uot \sum_{\gql} \frac{1}{N}\sum_{n'} \frac{\gL^{\gql}_{nn'\kk}}{\gee_{\nk}-\gee_{\npk}} \times (2 B(\goql) +1),
\label{eq:HAC_DW_correction}
\end{align}
where $\gL^{\gql}_{nn'\kk}$ is an expression written in terms of $\nabla \Vscf$ and obtained by imposing the translation invariance of $\gD \gee_{n\kk}$ when all atoms in the crystal are displaced of the same amount from their equilibrium positions~\cite{allen_heine1976}. It is worth noticing that from Eqs.~\eqref{eq:HAC_Fan_correction} and \eqref{eq:HAC_DW_correction} when the temperature ($T$) vanishes the energy correction does not vanish due to the $(2 B(\goql) +1)$ factor yielding the zero-point motion renormalization (ZPMR).

\section{Results and discussion}

\begin{figure*}
\includegraphics[width=1.5\columnwidth]{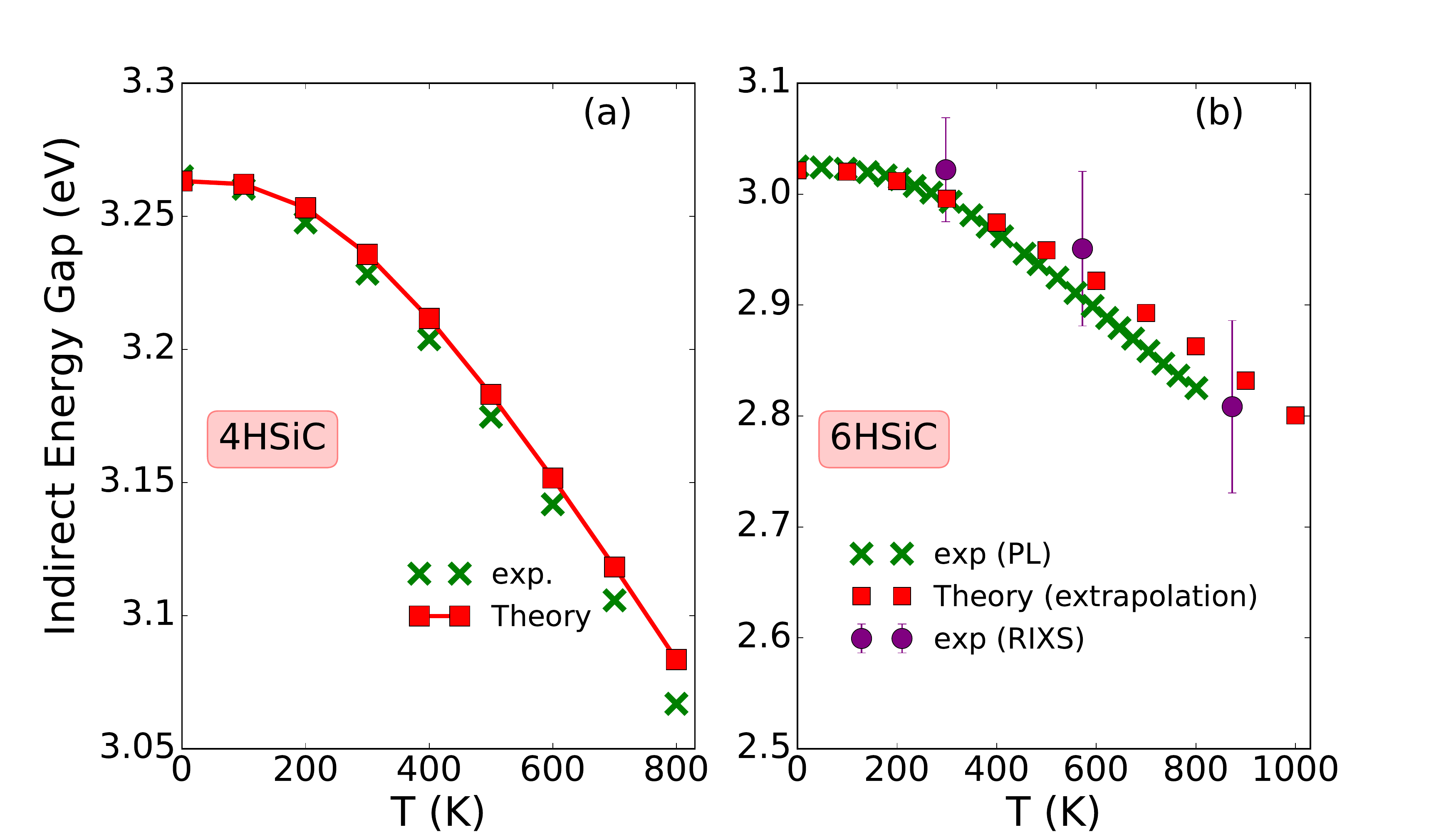}\\
\includegraphics[width=0.78\columnwidth]{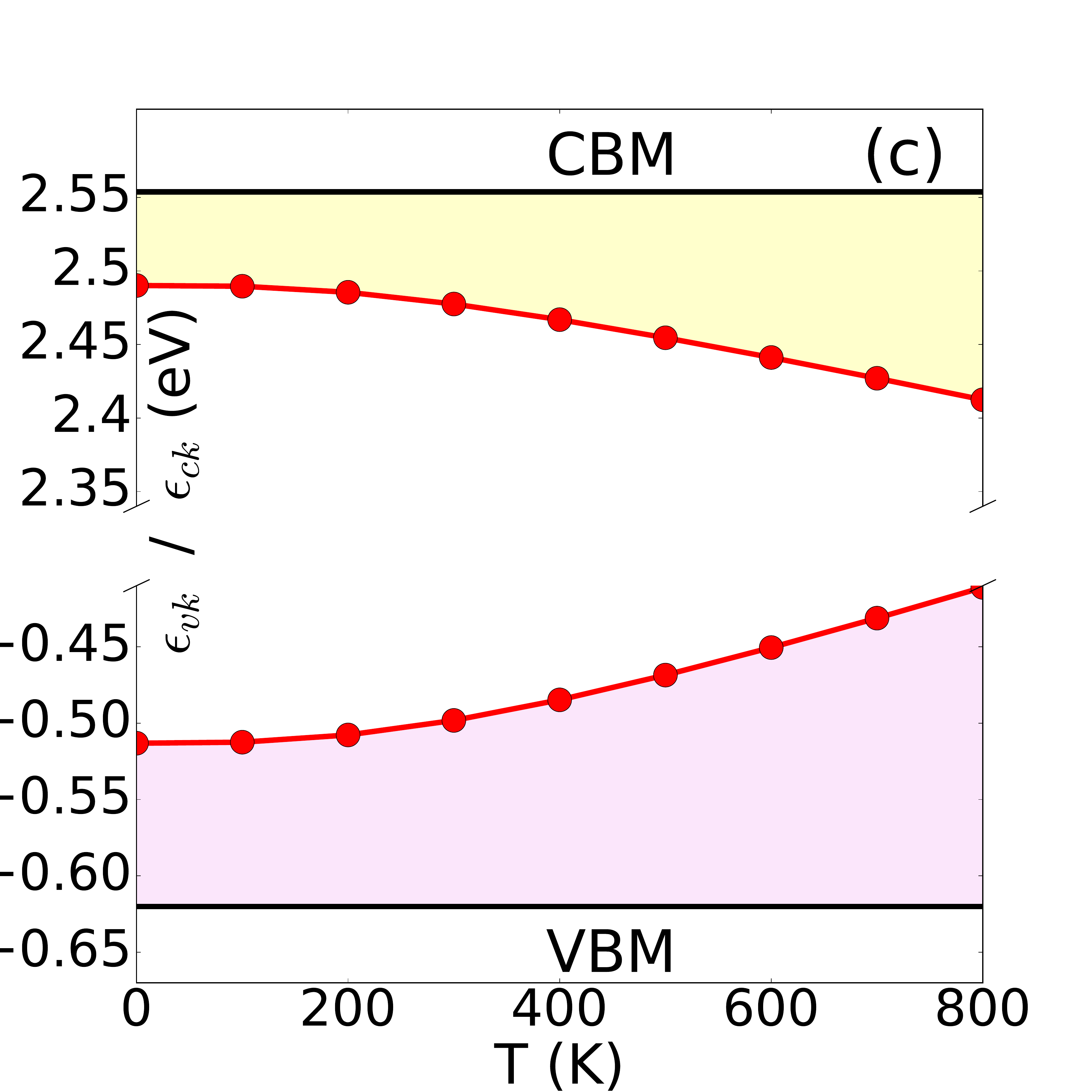}
\includegraphics[width=0.78\columnwidth]{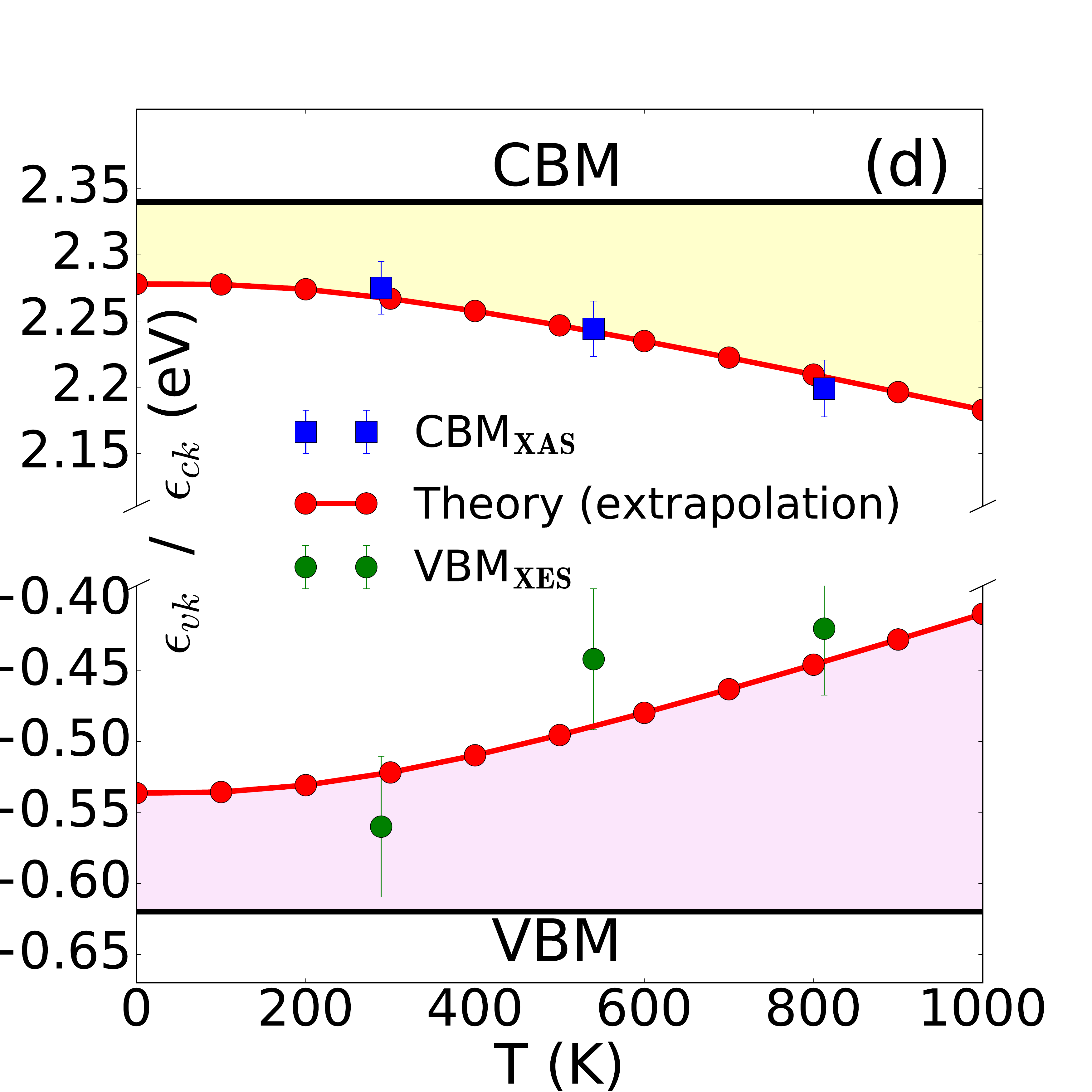}
\caption{\label{fig:bands_vs_T_4H6H} Calculated thermal evolution of indirect band gaps in the temperature range $0-800$~K for $4H$ SiC (a) and $0-1000$~K for $6H$ SiC (b). The calculated curves are aligned to the experimental data at $T=0~ \text{K}$ after having applied the GW correction to the indirect gap. For $6H$ an extrapolation at 150 bands (red squares: estimated convergent data) is given for comparison. The experimental data (green crosses in (a) and (b)) of the optical gaps are taken from Refs.~\onlinecite{Choyke1969, Pscajev2009}.  
The others (violet dots) have been extracted from RIXS spectra~(Ref.~\onlinecite{Miedema2014}), with correction of the core-hole exciton binding energy included. 
Thermal evolution of calculated VBM and CBM of (c) $4H$ SiC and (d) $6H$ SiC, where the latter is compared to the XAS and XES measurements (Ref.~\onlinecite{Miedema2014}) rescaled of 99.5~eV and 98.65~eV, respectively, in order to match the temperature evolution of CBM and VBM. Notice that on top of the electron-phonon correction the quasi-particle correction has been also added.  
}
\end{figure*}

Since the thermal expansion minutely increases the lattice constants of hexagonal SiC~\cite{Nakabayashi2006}, the thermal evolution of the band structure and of the electronic gap will be caused mainly by the electron-phonon interaction. We assess first the HAC approach to describe the thermal evolution of $4H$ and $6H$ SiC indirect band gaps [see Fig.~\ref{fig:SiC}(c) for a sketch of indirect band gaps in both samples]. At $T=0 \text{K}$, we found 0.17~eV and 0.14~eV for $4H$ and $6H$ SiC, respectively as of the indirect electronic band gaps. These values are about 5\% of the indirect electronic band gaps, revealing an intermediate ZPMR value in between bulk silicon and diamond~\cite{Monserrat2014,Cardona2005}. The calculated temperature dependence of the indirect band gap can be compared to that of the observed optical gap of SiC crystals~\cite{Choyke1969, Pscajev2009}. In Fig.~\ref{fig:bands_vs_T_4H6H}(a) and \ref{fig:bands_vs_T_4H6H}(b) the calculated temperature-dependent indirect band gaps are aligned to the experimental data at $T=0 \text{K}$ after having applied the electron quasiparticle (QP) correction as a simple scissor on the DFT-LDA band gaps. The temperature evolution agrees very well with the experimental data as derived from the optical gaps for a wide range of temperatures, which allows us to predict a shrinking of the electronic gap of about 0.35~eV ($4H$ SiC) and 0.30~eV ($6H$ SiC) at $T=800 \text{K}$. The shrinkings of the electronic gap of $6H$ SiC have been estimated as we will explain later and in SM in details. 

The above briefly mentioned opening of the indirect gap due to QP effects has been calculated adopting the Godby-Needs plasmon-pole model~\cite{godby1989PPA, GW1, GW2, Bockstedte2010} obtaining 0.92~eV QP-level correction for both $4H$ and $6H$ SiC. We used $18\times18\times6$ and $16\times16\times3$ k-point sets for $4H$ and $6H$ SiC, respectively, with 200 bands in the numerically convergent GW~\cite{GW1, GW2} calculations. Our correction results to be lower with respect to those calculated by Ummels et al.\cite{Ummels1998} [see Table~\ref{tab1:energies}]. This is due to the different plasmon pole model used [see Ref.~\cite{PPM-EF,CD,Stankowski} and SM for a further discussion on previous QP calculations on these systems]. 
Our LDA energies for the highest valence band and the lowest conduction bands at $\Gamma$ and $M$ points, in agreement with literature \cite{Kackell1994,park1994,Wenzien1995}, are listed in Table~\ref{tab1:energies} together with previous GW gaps~\cite{Ummels1998} and experimental data~\cite{Choyke1964}. The experimental data comes from optical measurements in which the exciton binding energies should be accounted for comparison to the calculated electronic band gaps. These binding energies are experimentally found to be 0.020~eV and 0.078~eV in $4H$ and $6H$ SiC~\cite{Sankin1975, Dubrovskii1975}, respectively, i.e., which are smaller than the expected accuracy of the GW calculations.

Next, we study the temperature dependence of the individual bands. In particular, we focus on the valence band maximum (VBM) and conduction band minimum (CBM) [see Figs.~\ref{fig:bands_vs_T_4H6H}(c) and (d)], whose difference provides the indirect band gap plotted in Figs.~\ref{fig:bands_vs_T_4H6H}(a) and (b), once it is opportunely rescaled in order to match the experimental data at T=0~K. Here we applied the  contribution of each band to the GW QP correction, -0.62 eV for the VBM and +0.30 eV for the CBM for both systems. Since VBM and CBM do not occur at the same point of the BZ, they have different symmetries and they interact differently with phonons. For $4H$ and $6H$ SiC, in fact, we predict an asymmetry in the band gap closing, where the contribution from the VBM is larger, respectively $\sim$63\% and $\sim$58\% of the total band gap shrinking. Here, we stress  that the latter result has been estimated. 
We used the converged data set for $4H$ SiC [see Fig.~S3 of SM] to extrapolate convergent data for $6H$ SiC because an explicit convergent calculation is computationally prohibitive as explained in SM.
In Fig.~\ref{fig:bands_vs_T_4H6H}(d) we report the temperature dependent XAS and XES spectra which depict respectively the CBM and VBM behaviour in $6H$ SiC~\cite{Miedema2014}. Both XAS and XES experimental data have been shifted to match the temperature evolution of CBM and VBM. We find a very good agreement with XAS derived data. The XES one also agrees with the estimated converged data set. Our theory well supports the observed temperature evolution of the indirect band gap of $4H$ SiC. 

\begin{figure*}
\includegraphics[width=0.49\textwidth]{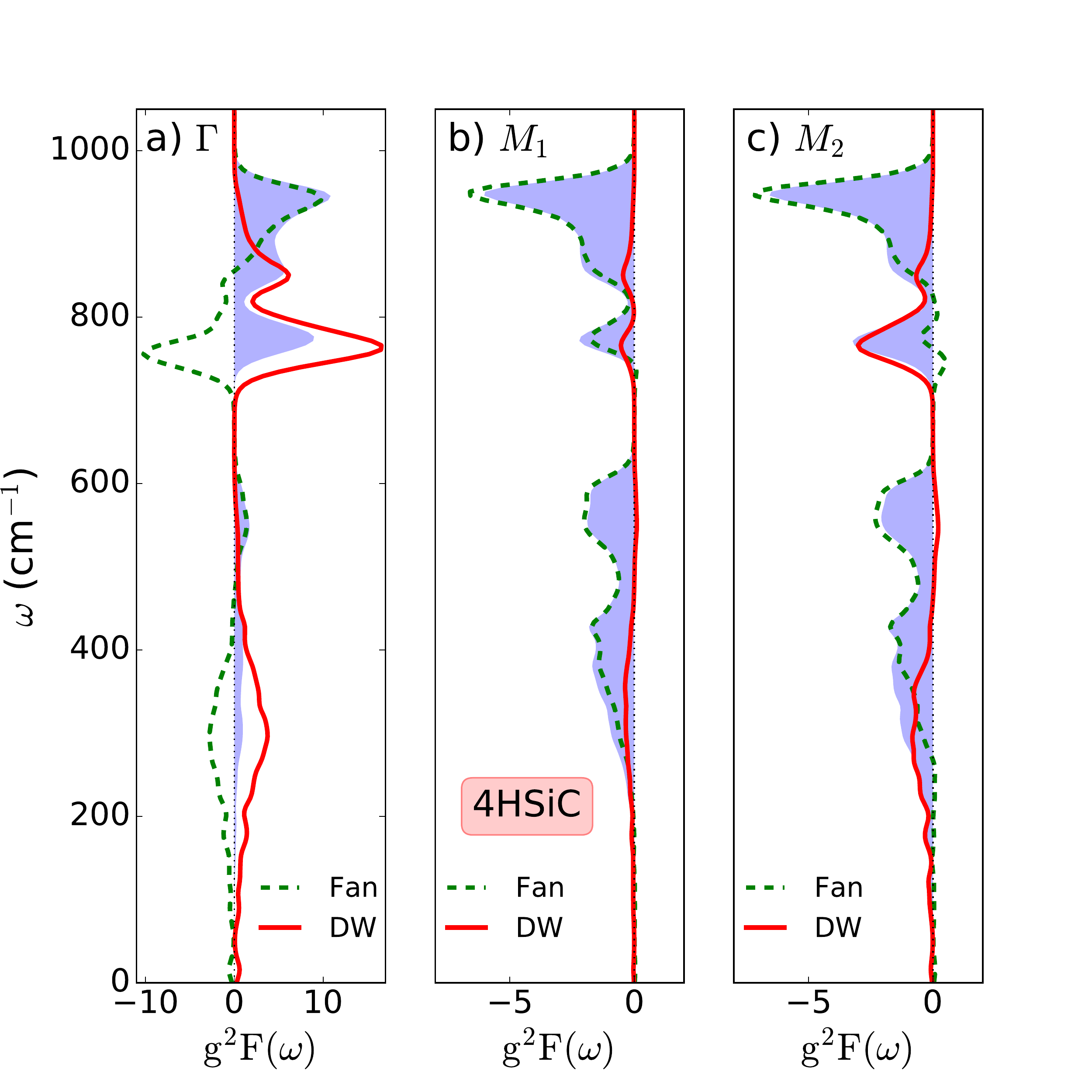}
\includegraphics[width=0.49\textwidth]{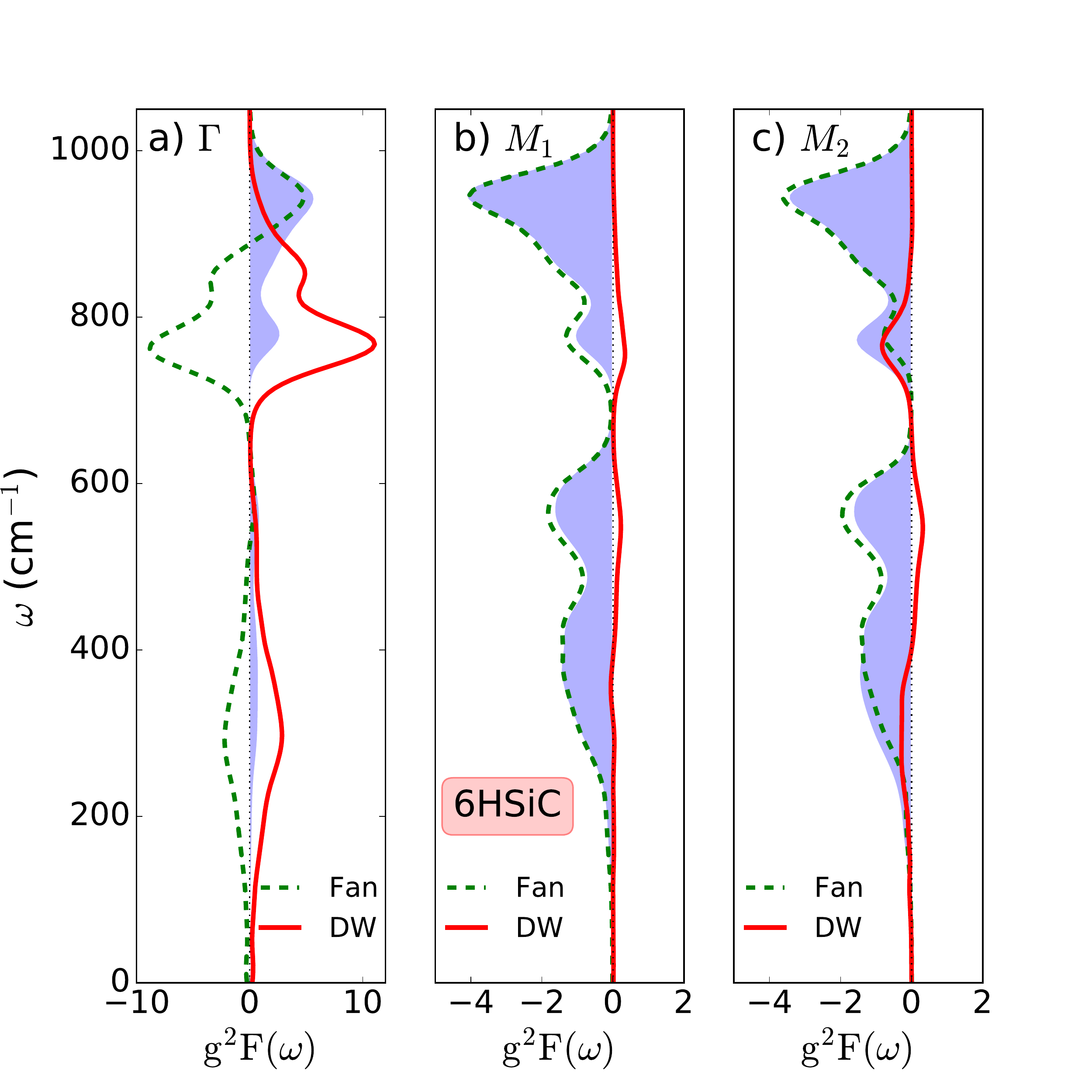}
\caption{\label{fig:Elfunc}Generalized electron-phonon Eliashberg function for $4H$ and $6H$ SiC (a) valence band maximum ($\Gamma$-point), (b) first conduction band minimum ($M_1$-point) and (c) second conduction band minimum ($M_2$-point). Fan and Debye-Waller (DW) contributions are shown separately with dashed and solid lines, respectively. The blue shaded areas represent the total Eliashberg function.}
\end{figure*}

The nature of the electron-phonon coupling induced renormalization in $4H$ and $6H$ SiC can be analyzed in the light of plotted Eliashberg function (Fig.~\ref{fig:Elfunc}), where the separate contribution of Fan and DW terms is highlighted. For both $4H$ and $6H$ SiC the VBM states are mostly coupled to the optical phonons (LO and TO) starting from 700~cm$^{-1}$. In this phonon frequency energy range both Fan and DW have a significant weight in the energy renormalization; while they almost cancel each other in the acoustic phonon frequency range. On the other hand, the CBM states (here called $M_1$) and the second lowest conduction band (here called $M_2$) also couple to acoustic phonons already from 220~cm$^{-1}$. Most of the contribution is given by the Fan term, which dominates all over the whole phonon frequency range. The difference between the CBM and VBM Eliashberg functions, being negative, is strictly related to the temperature dependence of the fundamental band gap shown in Fig.~\ref{fig:bands_vs_T_4H6H}. The square modulus of VBM wavefunction shows that the charge density is mainly localized on the Si-C bond of the hexagonal Si-C layer (off-axis Si-C bonds). The coupling with planar optic modes, as highlighted by the Eliashberg functions, is then justified, as these vibrations change the length of the interatomic bonds in which the charge density associated with the VBM resides~\cite{Monserrat2014}. The CBM states are localized in the interstitial places of $4H$ and $6H$ SiC lattice~\cite{Matsushita2012}, in contrast to the bond localized charge density of the VBM. The coupling with optical modes slightly looses weight and it is transferred to acoustic modes.

We observe now that the calculated lowest energy conduction bands of $4H$ SiC labeled as $M_1$ and $M_2$ are quite close in energy [see Tab.~\ref{tab1:energies}]. According to a recent measurement, the energy separation of the two lowest conduction bands at $M$ point of the Brillouin-zone is 144$\pm$2~meV at 2~K~\cite{Klahold2017}. If these two bands crossed at elevated temperatures then this could seriously affect the $n$-type conductivity of the SiC based electronic devices. We find that the energy separation between $M_1$ and $M_2$ only reduces by 5~meV ($<5\%$) going from 0~K to 800~K temperatures [see Fig.~S4 in SM] because both $M_1$ and $M_2$ bands shift downwards with increasing temperatures. We note that similar properties are found in $6H$ SiC [see Fig.~S4 in SM]. The calculated effective masses of the $M_1$ and $M_2$ bands [see Table~SII in SM and Ref.~\cite{Son1995}] imply that the conductivity of the $M_2$ band is smaller than that of the $M_1$ band except in the $K-M$ direction. However, the overall conductivity~\cite{GrossoPastori,Fanfoni} still increases [see Fig.~S5 in SM] because of a larger number of typical nitrogen donors ionization at $k$ and $h$ sites at 120 and 60~meV~\cite{Takase2013}, respectively, with increasing temperatures. This paves the way for space agencies to employ $4H$ SiC based integrated circuits to probe the surface of Venus, where the temperature is $\approx$730~K~\cite{Neudeck2016}.    

Our fundamental study has implications on the quantum bits hosted by $4H$ SiC such as neutral divacancies~\cite{Gali:PSSB2011, Koehl11}. Divacancy qubits are initialized and readout by optically detected magnetic resonance (ODMR)~\cite{Koehl11}, where illumination at about 1.17~eV used in this process may lead to a dark state, i.e., a permanent loss of qubits~\cite{Wolfowicz2017, Golter2017, Magnusson2018}. The qubit state can be restored by applying $\sim$1.3~eV optical excitation~\cite{Wolfowicz2017, Golter2017, Magnusson2018}, where the nature of the dark state was debated in the literature~\cite{Wolfowicz2017, Golter2017, Beste2018}.  According to one of the most recent studies~\cite{Magnusson2018}, the dark state can be identified as the negative charge state of divacancies where optical excitation at $\sim$1.25~eV is the threshold of photoionization of the electron from the in-gap defect level to the conduction band edge at cryogenic temperatures. The coherent control of defect spins in $4H$ SiC can be achieved even at high temperatures up to 600~K~\cite{Yan2018} paving the way for SiC-based broad-temperature-range quantum sensing such as magnetic and temperature sensing. As the typical excitation energy of the qubit and the threshold for photoionization of the dark state are close in energy, the temperature dependent band edge and photoionization energies can seriously affect the stability of the charge state of divacancies, i.e., the operation of divacancy qubits. We find that the CBM of $4H$ SiC shifts down by 0.1~eV going from cryogenic temperatures up to 600~K. This implies that the threshold for photoionization reduces $\sim$0.1~eV. We conclude that high temperature operation of $4H$ SiC divacancy qubits would be stable against photoionization.

\section{Summary \& Conclusion}
In this study, we applied density functional theory on the electronic structure and phonons in $4H$ and $6H$ SiC, and determine the electron-phonon coupling within Heine-Allen-Cardona (HAC) approach\cite{allen_heine1976,allen_cardona_1981,allen_cardona_1983}. We find a sizable temperature dependent renormalization energy on the electronic structure for both crystals. 
As a consequence, both valence and conduction band edges shift with temperature thus affecting the operation of defect qubits at elevated temperatures. We predict that the operation of divacancy qubits in $4H$ SiC~\cite{Gali:PSSB2011, Koehl11} is stabilized at elevated temperatures against bleaching caused by photoionization. Furthermore, our \emph{ab initio} results indicate that the two lowest conduction bands do not cross at elevated temperature and the conductivity of doped $4H$ SiC is not much affected, thus SiC electronics conforms to high temperature operation.   

We conclude that SiC exhibits favorable properties for hosting electronic devices at extreme high temperatures operation which is important for next generation sensors and electronics in space missions. Our study has an impact on quantum technology applications too, and serves as a template for similar studies in other semiconductors.

\section*{Acknowledgement}
E.C.\ acknowledges support by the Programma per Giovani Ricercatori - 2014 ``Rita Levi Montalcini'', French HPC computational resources from GENCI-CCRT and TGCC, through the Grant No.\ 2017-910165 and ISCRAC Italian Cineca project No.  HP10CEPHVH.
The support from the EU Commission and the National Office of Research, Development, and Innovation in Hungary (NKFIH) in the QuantERA Nanospin project (NKFIH Grant No.~127902), the support from NKFIH in the National Quantum Technology Program (NKFIH Grant No.~2017-1.2.1-NKP-2017-00001) and the National Excellence Program (Grant No.~KKP129866) is acknowledged.



%

\end{document}


\title{Supplemental Materials\\for the paper entitled\\Thermal evolution of silicon carbide electronic bands}
\author{E. Cannuccia}
\affiliation{Aix-Marseille Universit\'e , Laboratoire de Physique des Interactions Ioniques et Moléculaires (PIIM), UMR CNRS 7345, F-13397 Marseille, France}
\affiliation{Dipartimento di Fisica, Universit\`a di Roma ``Tor Vergata'', Via della Ricerca Scientifica 1,I--00133 Roma, Italy}
\author{A. Gali}
\affiliation{Wigner Research Centre for Physics, PO.\ Box 49, H-1525, Budapest, Hungary}
\affiliation{Department of Atomic Physics, Budapest University of Technology and Economics, Budafoki \'ut 8., H-1111 Budapest, Hungary}

\maketitle
 
\section{Description of the hexagonal silicon carbide crystals} 

Silicon carbide (SiC) has over 250 polytypes which share the same hexogonal lattice in the basal plane and different stacking sequences of Si-C bilayers perpendicular to the basal plane. The most important polytypes are the so-called $4H$ and $6H$ SiC as shown in Fig.~\ref{fig:SiC}(a) and Fig.1(a) in the main text. They consist of 4 and 6 Si-C double layers having a close-packed hexagonal arrangement. There are three types of hexagonal close packing (A, B and C) in arranging the Si-C double atomic layers.  In $4H$ and $6H$ the stacking sequences are ABCB$|$ABCB$|\dots$ and ABCACB$|$ABCACB$|\dots$, respectively. Each atom (Si or C) has a local quasicubic ($k$) or hexagonal ($h$) environment with respect to the immediate neighbors. The experimental lattice constants are {\textbf a} = 3.08~\AA\, with {\textbf c} = 10.08~\AA\, \cite{Stockmeier2009} and {\textbf a} = 3.08~\AA\, with {\textbf c} = 15.12~\AA\, \cite{Stockmeier2009} for $4H$ and $6H$ SiC, respectively, and the ideal values for the internal parameters follow from the space group symmetry $C^{4}_{6v}$\cite{Kackell1994}. 
 
\section{Electronic structure of hexagonal silicon carbide crystals}

We applied density functional theory (DFT) within local density approximation (LDA) as reported in the main text. The calculation of the phonon dispersion curves required the optimization of the lattice constants. This results into {\textbf a} = 3.07~\AA\, with {\textbf c} = 10.0~\AA\, \cite{Stockmeier2009} and {\textbf a} = 3.07~\AA\, with {\textbf c} = 15.0~\AA\, for $4H$ and $6H$ SiC, respectively, that are close to the experimental values. The optimized internal parameters remain close to the ideal ones \cite{Kackell1994}.

\begin{figure}[hb]
\includegraphics[width=0.2\textwidth]{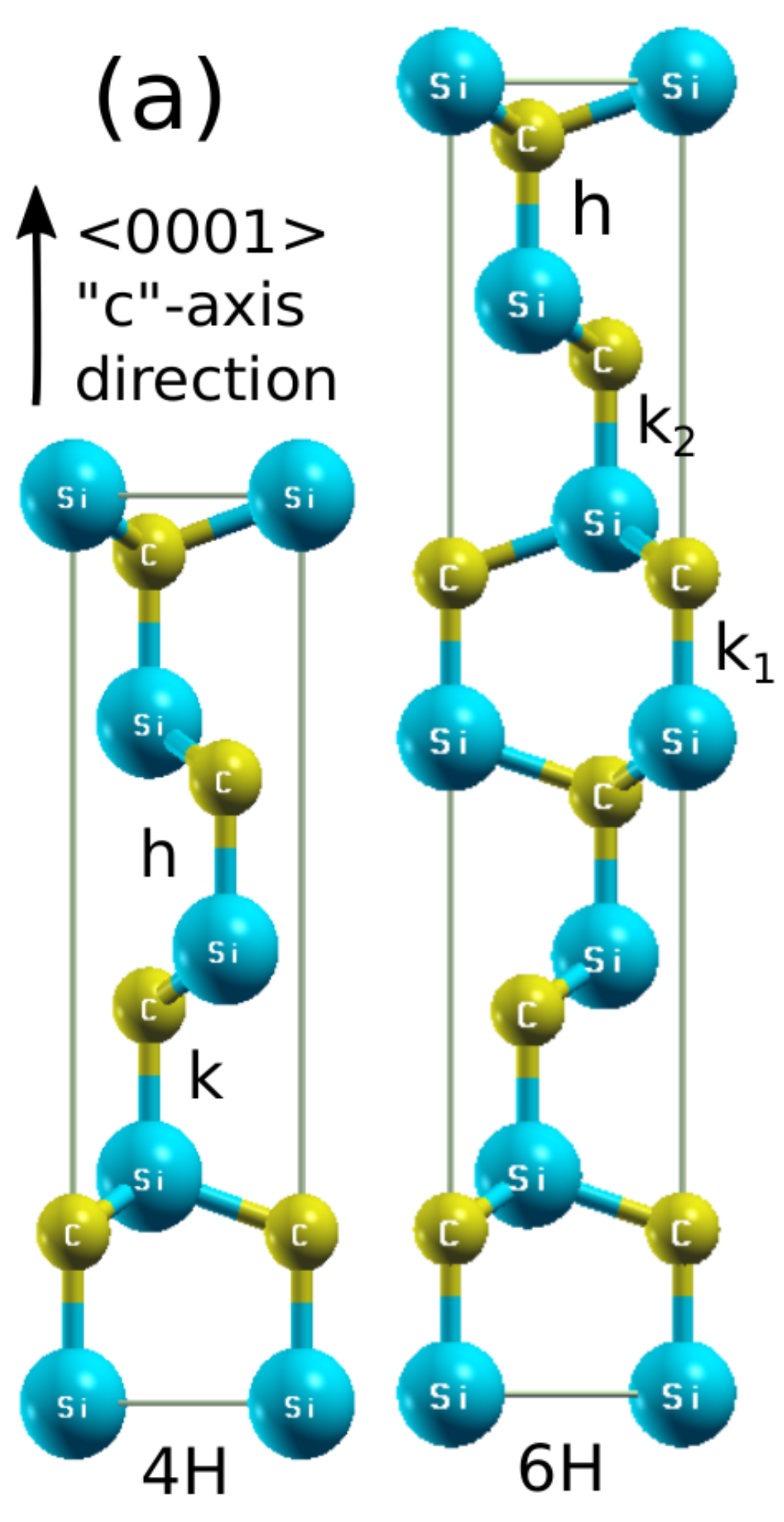}
\includegraphics[width=0.22\textwidth]{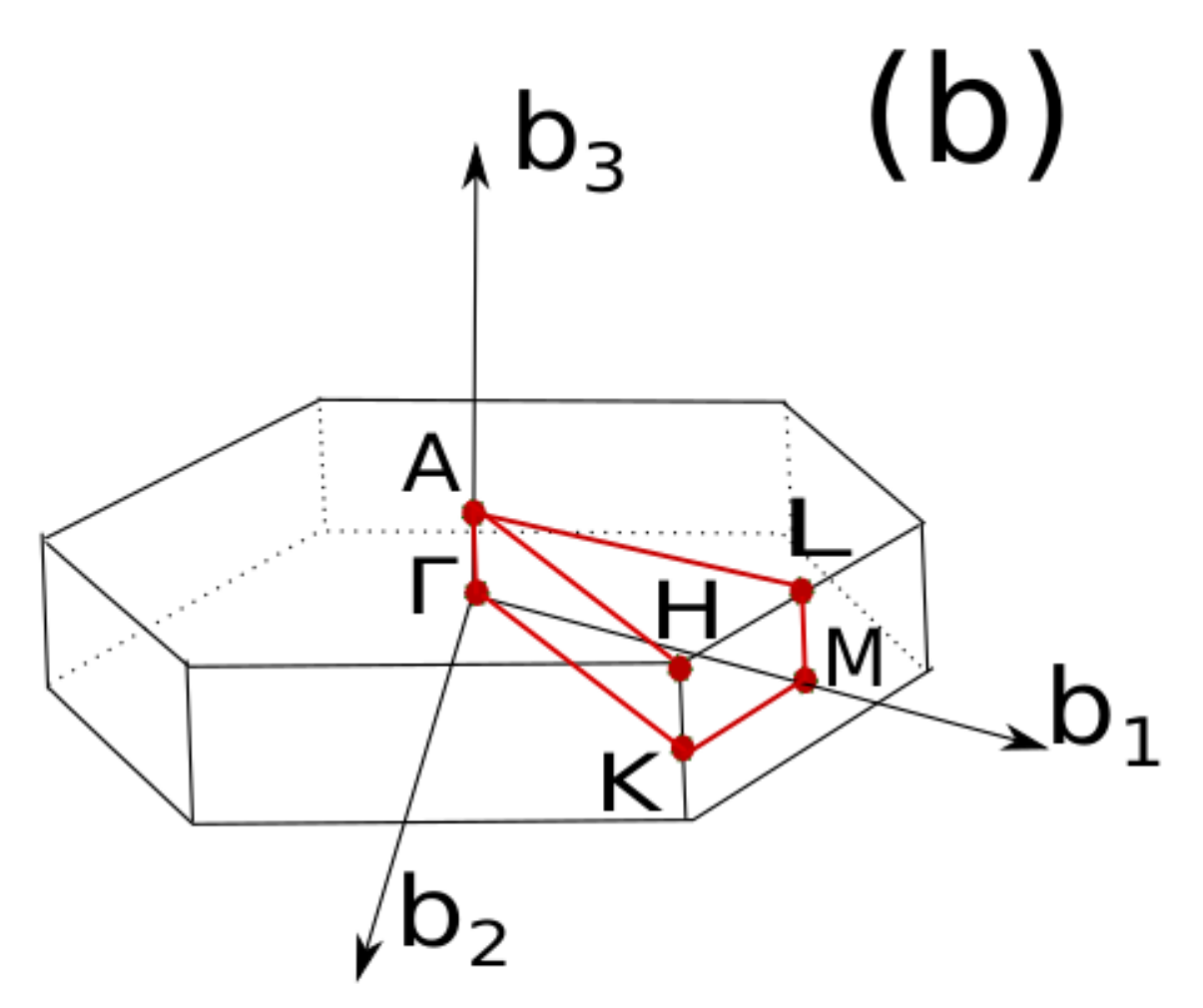}
\includegraphics[width=0.45\textwidth]{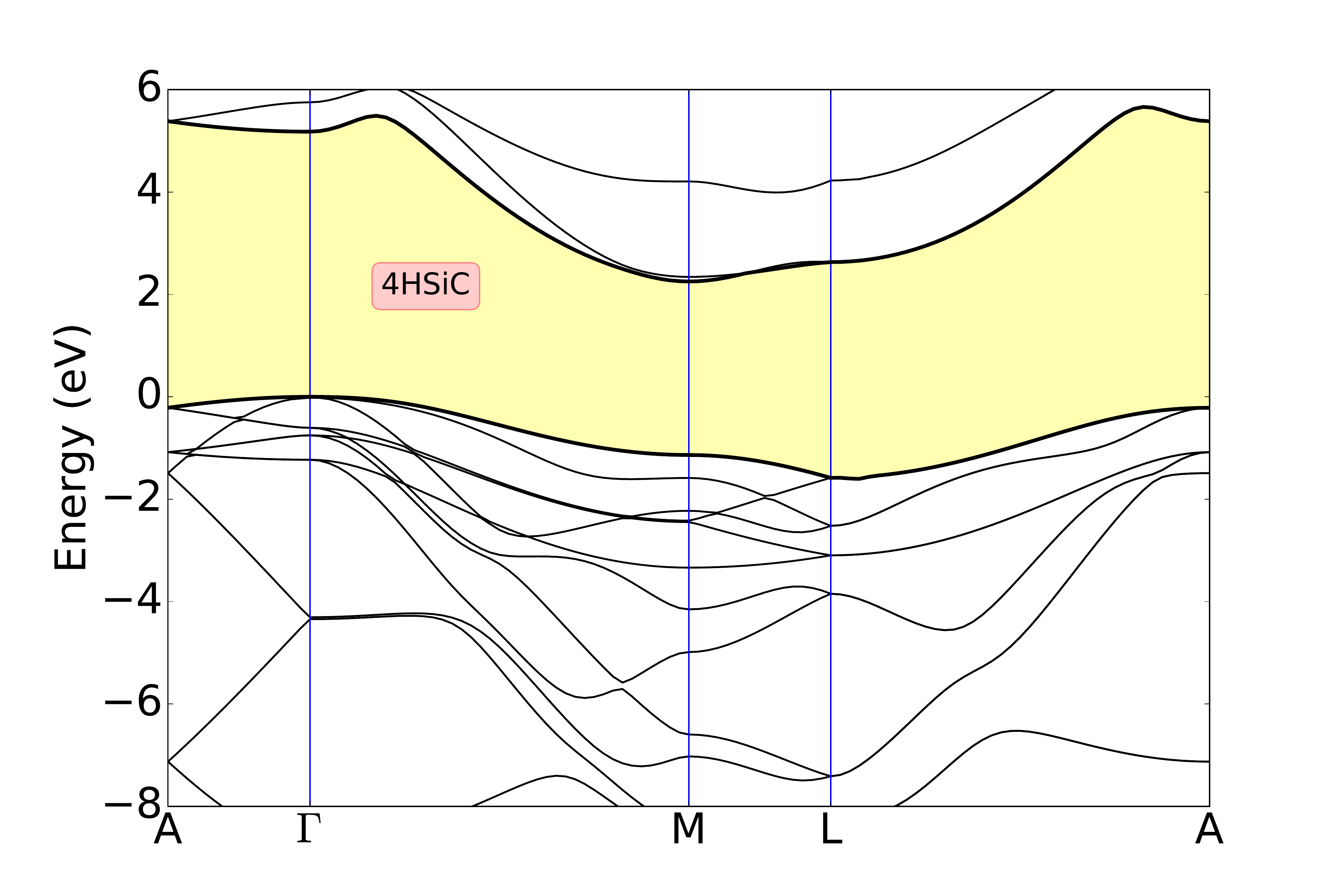}
\includegraphics[width=0.45\textwidth]{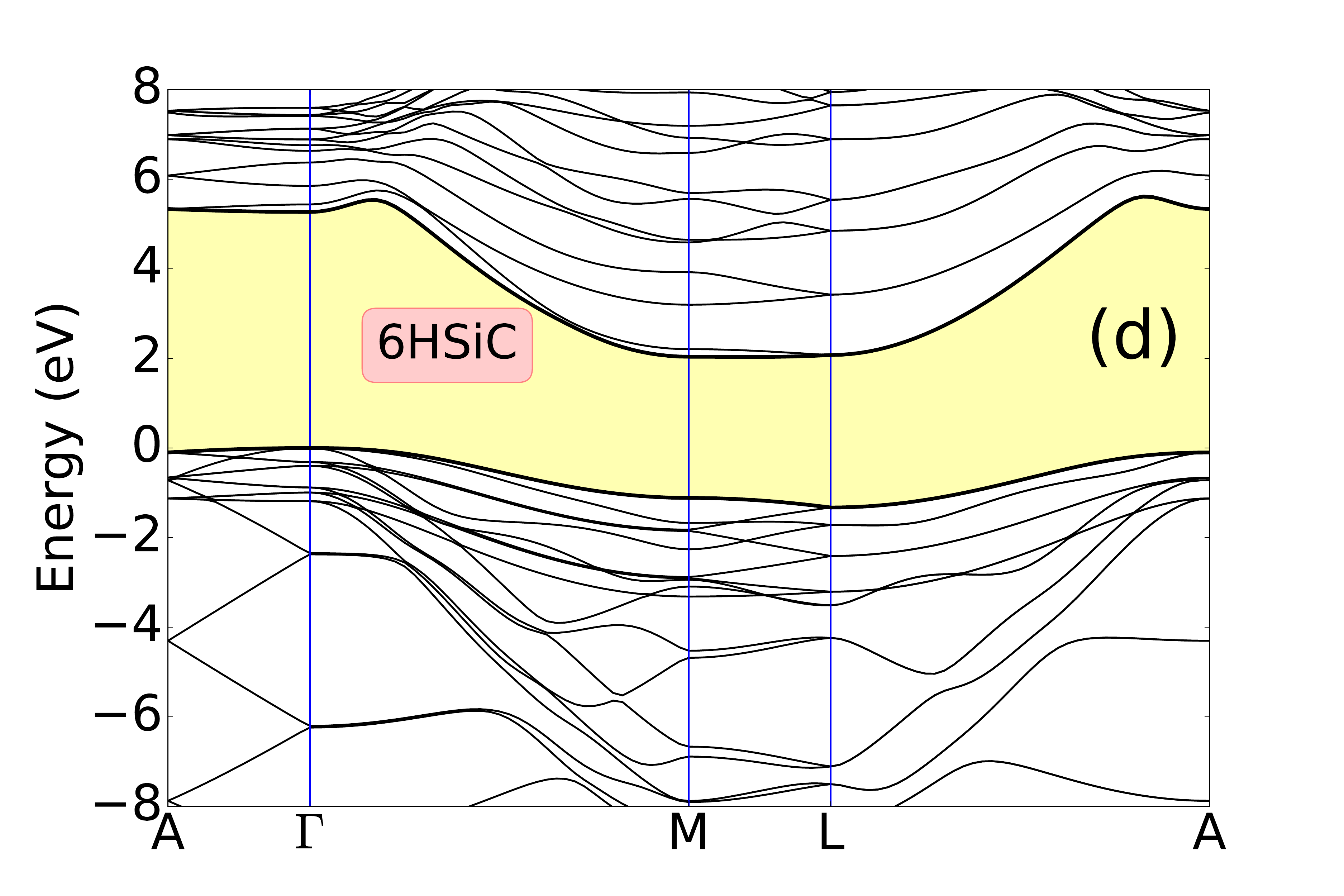}
\caption{\label{fig:SiC}(a) Primitive cells of $4H$ and $6H$ SiC with $k$ ($k_1$, $k_2$) and $h$ Si-C bilayers where $k$ and $h$ refer to quasicubic and hexagonal sites. (b) Brillouin-zone of hexagonal SiC polytypes. (c) Calculated LDA band structure of $4H$ SiC. The valence band maximum (VBM) is at the $\Gamma$ point, whereas the conduction band minimum (CBM) is at the $M$ point. (d) Calculated LDA band structure of $6H$ SiC. The firt conduction band is very flat along $M-L$ line~\cite{Kackell1994, Lambrecht2001}. While the VBM is still at $\Gamma$ point, we find the CBM in between $M$ and $L$. The VBM energies are aligned to zero, and the band gap is highlighted by yellow color (shadowed).} 
\end{figure}
 
$4H$ and $6H$ SiC are indirect semiconductors with relatively large band gaps at room temperature, $3.27$ and $3.02$ eV, respectively~\cite{Landolt-Bornstein}. The calculated electronic band structures are shown in Fig.~\ref{fig:SiC}(c) and (d) versus high symmetry lines $A$-$\Gamma$-$M$-$L$-$A$ within the Brillouin-zone (BZ) of the hexagonal system (see Fig.\ref{fig:SiC}(b)). The overall features show the agreement with previous calculations \cite{Kackell1994,park1994,Wenzien1995}. The top of the occupied valence-band states is located at $\Gamma$ in both polytypes. As reported by \cite{Kackell1994,Lambrecht2001} the exact position of conduction band minimum depends on the details of the calculation, on the ratio $c/a$ of the hexagonal lattice constants, as well as on the atomic positions inside the hexagonal unit cell. We find the conduction band minimum at $M$ for $4H$ SiC and at about $0.5$ $M-L$ for $6H$ SiC. However, the lowest conduction band is rather flat between $L$ and $M$ in $6H$ SiC case. For this reason, we assume that the bottom of the conduction band occurs at the $M$ point of the BZ, yielding just a minor error in the calculated LDA fundamental gap for $6H$ SiC. Moreover, we observe that at the CBM two bands are very close in energy. We will label them in the follow as $M_1$ and $M_2$.
 
At DFT-LDA level we obtain 2.25~eV for $4H$ SiC and 2.04~eV for $6H$ SiC, about 1.0~eV lower than the corresponding measured values, and in agreement with other theoretical works~\cite{Wenzien1995, Kackell1994}. The discrepancy in the absolute value of the indirect band gaps comes from the well-known self-interaction error of the LDA but the curvature of the individual bands are well reproduced by this approximation. As a consequence, the effective masses of the electrons, i.e., the inverse of the second derivatives of the energies around the $M$ point for the conduction band minimum, can be accurately calculated by DFT-LDA.

\section{Quasiparticle correction in the electronic structure of hexagonal silicon carbides} 

Assuming DFT-LDA orbitals as a good approximation to the quasi-particle (QP) orbitals, the first-order
correction to the Kohn-Sham eigenenergies is obtained within a single iteration of Hedin's GW approximation~\cite{GW1, GW2}. For $6H$ SiC, the exact position of the conduction band minimum (CBM) on the $L-M$ line has been under discussion~\cite{park1994, Kackell1994, Lambrecht2001}. Wenzien \emph{et al.}~\cite{Wenzien1995} found a tendency of CBM of $6H$-SiC to move towards $M$ point after the inclusion of QP effects. They used a simplified GW method, meaning a model dielectric function and an approximate treatment of the local filed effects. Ummels \emph{et al.}~\cite{Ummels1998} performed GW calculations for the $2H$, $4H$, and $6H$ SiC polytypes using the Engel and Farid\cite{PPM-EF} plasmon pole model for the description of the energy dependence of the screened interaction. They obtained G$_0$W$_0$ indirect band gaps of $3.35$ eV and $3.24$ eV for $4H$ and $6H$ SiC, respectively (see Table I in the main manuscript). 
Our correction obtained with Godby-Needs plasmon-pole model~\cite{godby1989PPA, GW1, GW2, Bockstedte2010} as implemented in {\tt Yambo}~\cite{Sangalli_2019} code, results to be lower with respect to those calculated by Ummels et al.\cite{Ummels1998}. This is due to the different plasmon pole model (PPM) used. Godby-Needs PPM best reproduces the model-free contour deformation (CD) results\cite{CD}, when other PP models are applied the gap obtained is significantly higher and in particular Engel and Farid PPM gives for ZnO a GW correction $0.3$-$0.4$ eV larger than the CD one~\cite{Stankowski}.

In our study, we have to rely on the DFT-LDA geometries as we have to calculate phonon bands which assumes to stay close to the global energy minimum of the adiabatic potential energy surface. Our GW calculations justify to apply a simple scissor correction on the DFT-LDA band gaps, in order to reproduce the experimental ones. 

\section{Phonons of hexagonal silicon carbides}
In view of the calculation of the temperature dependent renormalization of the electronic gap, the phonon dispersion curve is an essential input for phonon-related properties. Besides the fact that the phonon dispersion curve reflects the structure of a material and its symmetries, phonon spectra are one of the key ingredients that help in understanding the coupling between electrons and lattice vibrations. Symmetry analysis shows that phonon modes can be decomposed into $N(A_1 \oplus B_1 \oplus E_1 \oplus E_2)$ modes, where $N=4, 6$ for $4H$ SiC and $6H$ SiC, respectively. Whose modes $A_1 \oplus E_1$ are acoustic and $(N-1) A_1 \oplus N B_1 \oplus (N-1) E_1 \oplus N E_2$ are optic modes. Except for $B_1$ modes which are infrared and Raman forbidden, all the other are Raman active modes. Out of them $A_1$ and $E_1$ modes are also infrared active. 

\begin{figure}
\includegraphics[width=0.45\textwidth]{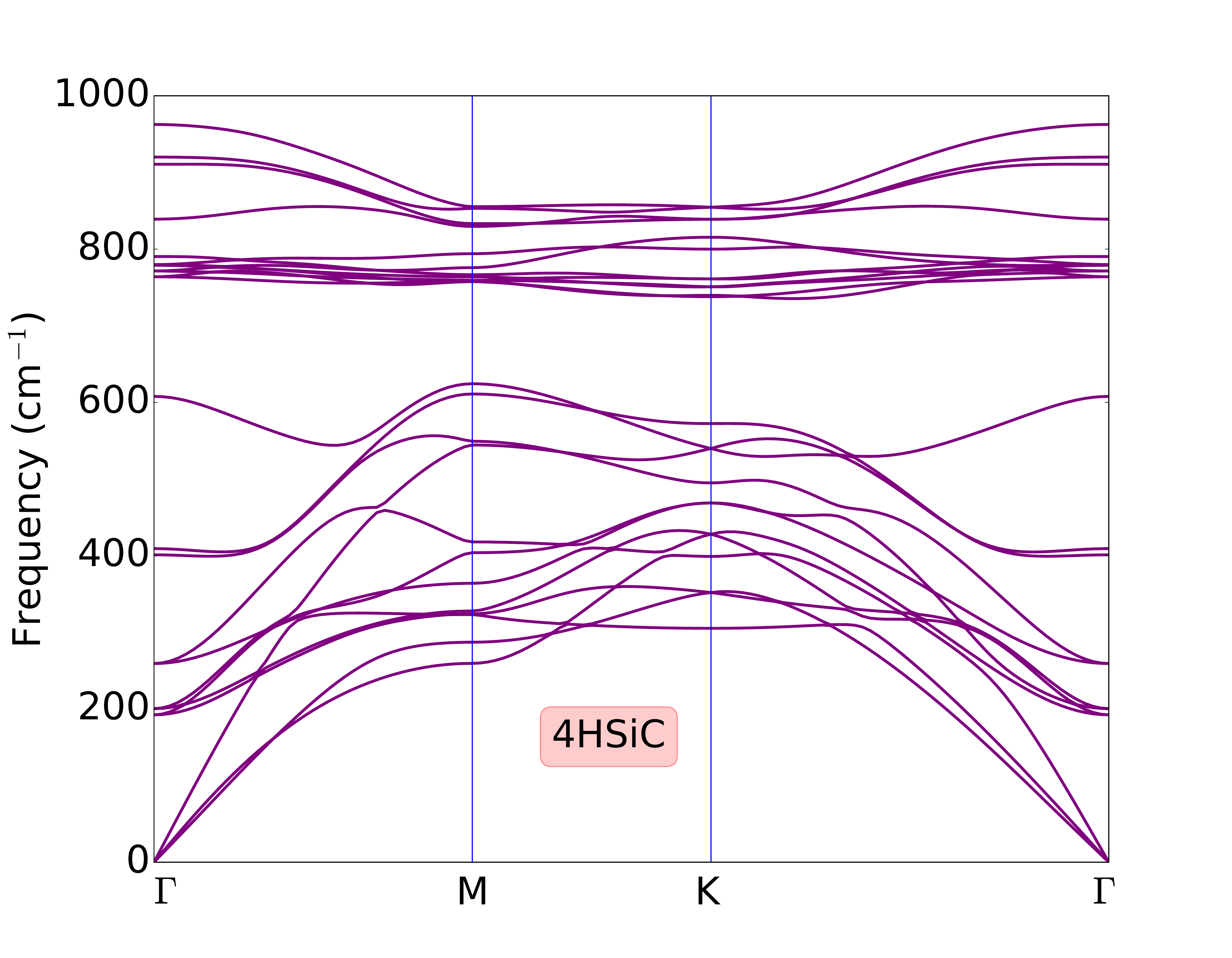}
\includegraphics[width=0.45\textwidth]{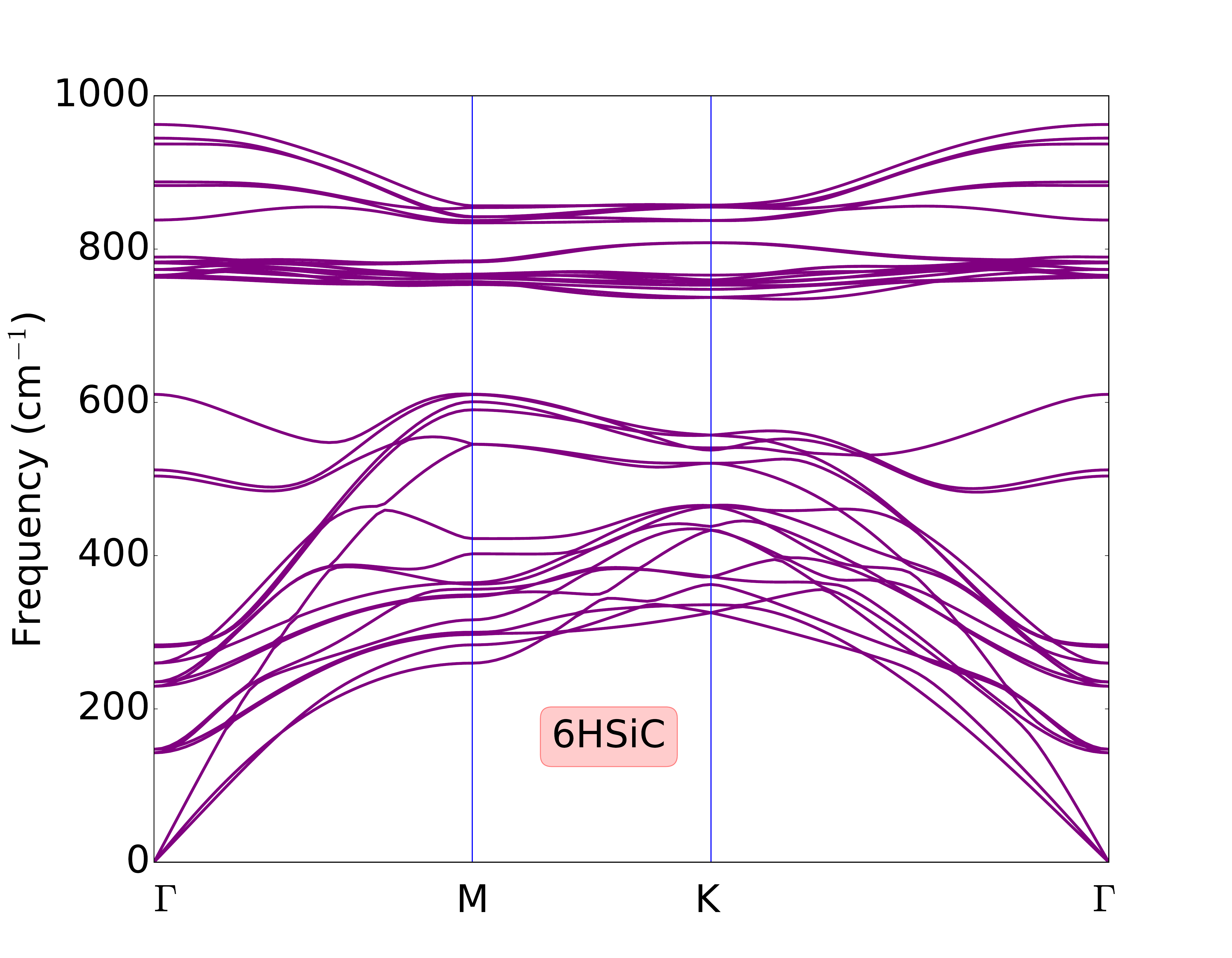}
\caption{\label{fig:SiC_PHdisp}Phonon band dispersion of $4H$ and $6H$ SiC as obtained within Density Functional Perturbation Theory (DFPT).}
\end{figure}
The calculated phonon dispersion curves are shown in Fig.~\ref{fig:SiC_PHdisp}. Table~\ref{tab2:4H_PHGamma} reports the theoretical and experimental phonon frequencies at $\Gamma$ point for $4H$ and $6H$ SiC. All the available experimental modes were readily assigned. The phonon modes with longer wavelength are divided into axial or basal plane modes, depending if the atoms displace parallel or perpendicular to the $c$-axis, as explained in Ref.~\cite{Patrick1968}. The planar modes consist of Raman and IR active $E_1$ modes and Raman active $E_2$ modes. The axial modes consist of Raman active $A_1$ and Raman inactive $B_1$ modes. Our vibrational frequencies at $\Gamma$ point are in excellent agreement with previously published calculations and with experimental phonon dispersion curves obtained by infrared and Raman spectroscopies~\cite{Feldman1968,Feldman1968bis,Nienhaus1995,Nakashima1997,Bluet1999}, being the average error smaller than 4~cm$^{-1}$. This supports our choice of using DFT-LDA eigenergies and DFPT-LDA phonon frequencies as key ingredients for the calculation of the electron-phonon interaction.   

\begin{table}[h]
\caption{\label{tab2:4H_PHGamma}$4H$- and $6H$-SiC optic (Opt.) and low energy acoustic (Ac.) phonon frequencies at $\Gamma$ point (in cm$^{-1}$) either in the basal plane (B.) or along c-axis (Ax.) and their corresponding irreducible representation (Irrep). $B_1$ modes are Raman and infrared (IR) forbidden. There are modes for both compounds that were not observed (N.O.).}
\begin{ruledtabular}
  \begin{tabular}{ccccc}                         
    Irrep & DFPT-LDA   & Exp.\footnote{Refs.~\onlinecite{Feldman1968,Nakashima1997}} & Exp.\footnote{Ref.~\onlinecite{Bluet1999}} & Branch  \\
          &  freq.   & Raman                                &  IR  reflectivity    &         \\\hline
    $E_2$ &  192.3   &  196.0                               &  -                   & B.Ac.\\
    $E_2$ &  200.3   &  204.0 		                        &  -                   & B.Ac.\\
    $E_1$ &  259.2   &  266.0                               &                      & B.Ac.\\
    $B_1$ &  401.1   &  -                                   &  -                   & Ax.Ac. \\
    $B_1$ &  409.2   &  -                                   &  -                   & Ax.Ac. \\
    $A_1$ &  608     & 610.0                                &  610.5               & Ax.Ac. \\
    $E_1$ &  763.9   & N.O.                                  &                      & B.Opt.\\
    $E_2$ &  771.4   & N.O.	                                &  -                   & B.Opt.\\
    $E_2$ &  779.7   & 776.0   		                        &  -                   & B.Opt. \\
    $E_1$ &  790.2   & 797                                  &                      & B.Opt. \\
    $A_1$ &  838.7   & 838                                  &  839                 & Ax.Opt.\\
    $B_1$ &  910.6   & -      		                        &  -                   & Ax.Opt.\\
    $B_1$ &  920.1   &  -      		                        &  -                   & Ax.Opt.\\  
    $A_1$ &  958.8   & 964                                  &                      & Ax.Opt. \\
  \end{tabular}
\end{ruledtabular}

\begin{ruledtabular}
  \begin{tabular}{ccccc}                         
    Irrep & DFPT-LDA   & Exp.\footnote{Ref.~\onlinecite{Feldman1968bis}} & Exp.\footnote{Ref.~\onlinecite{Bluet1999}} & Branch  \\
          &  freq.   & Raman                                &  IR reflectivity     &         \\\hline
    $E_2$ &  142.7   &  145.0                               &  -                   & B.Ac.\\
    $E_2$ &  147.4   &  149.0 		                        &  -                   & B.Ac.\\
    $E_1$ &  229.6   &  236.0                               &                      & B.Ac.\\
    $E_1$ &  235.2   &  241.0                               &                      & B.Ac.\\
    $E_2$ &  259.7   &  262.0 		                        &  -                   & B.Ac.\\
    $B_1$ &  281.0   &  -                                   &  -                   & Ax.Ac. \\
    $B_1$ &  283.4   &  -                                   &  -                   & Ax.Ac. \\
    $A_1$ &  504.1   & 504                                  &                      & Ax.Ac. \\
    $A_1$ &  511.9   & 508                                  &                      & Ax.Ac. \\
    $B_1$ &  610.5   &  -                                   &  -                   & Ax.Ac. \\
    $E_2$ &  763.1   & 766	                                &  -                   & B.Opt.\\
    $E_1$ &  765.7   & 769                                 &                       & B.Opt.\\
    $E_1$ &  773.3   & 777                                 &                       & B.Opt.\\
    $E_2$ &  782.0   & N.O.	                                &  -                   & B.Opt.\\
    $E_2$ &  783.2   & 788   		                        &  -                   & B.Opt. \\
    $E_1$ &  789.5   & 797                                  &  789                 & B.Opt. \\
    $B_1$ &  837.9   &  -      		                        &  -                   & Ax.Opt.\\
    $A_1$ &  882.9   & N.O.                                 &  884                 & Ax.Opt.\\
    $A_1$ &  887.6   & 889                                  &  889                 & Ax.Opt.\\
    $B_1$ &  937.1   &  -      		                        &  -                   & Ax.Opt.\\
    $B_1$ &  944.8   &  -      		                        &  -                   & Ax.Opt.\\  
    $A_1$ &  960.1   & 964                                  &                      & Ax.Opt. \\
  \end{tabular}
\end{ruledtabular}
\end{table}

\section{Additional data on the calculated electron-phonon interaction}
This section provides details about the convergence of the temperature dependent electron-phonon coupling of the valence bands as well as the the calculated temperature shifts for the two lowest energy conduction bands around $M$ point.

\begin{figure}[h]
\includegraphics[width=0.7\textwidth]{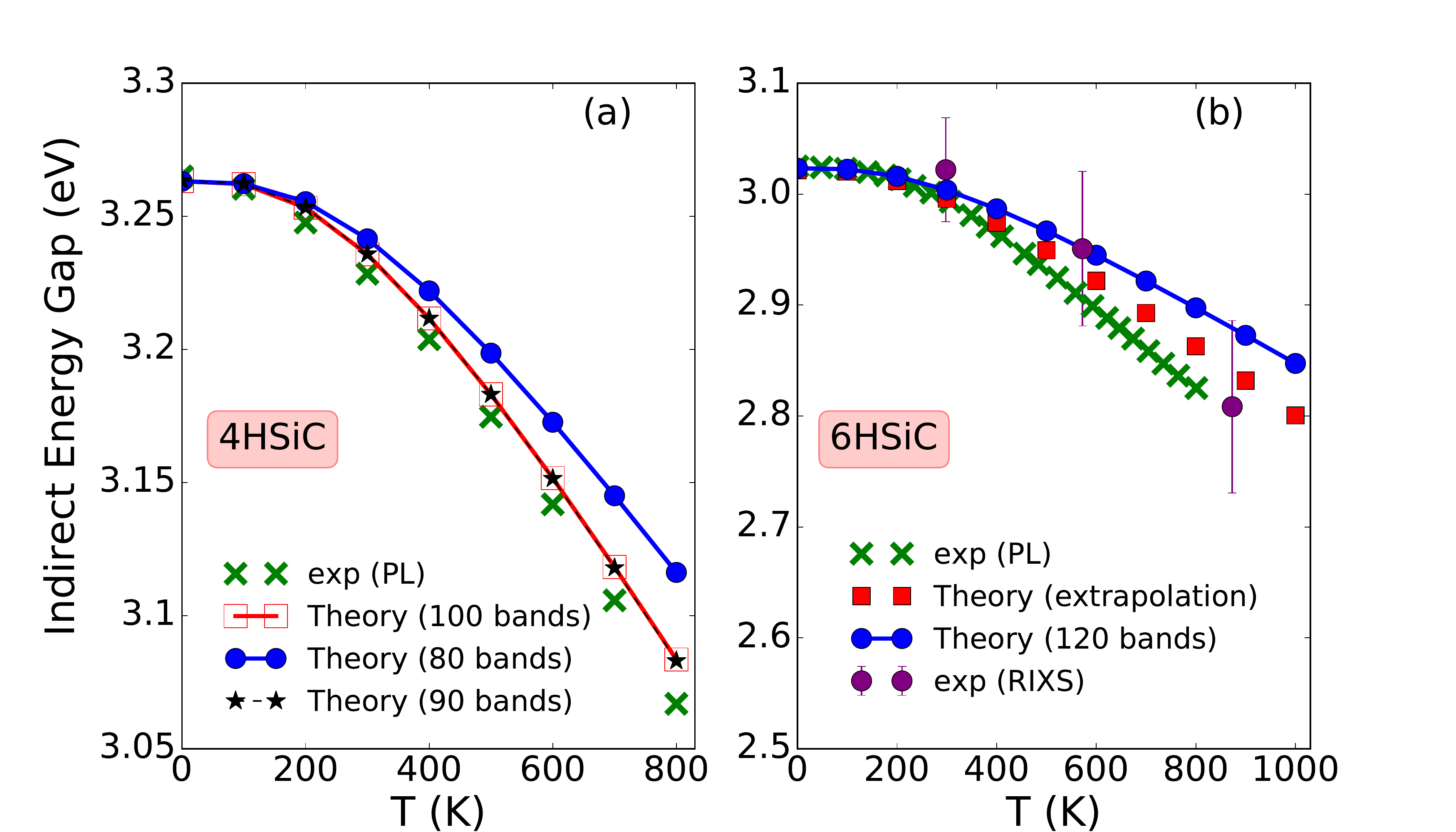}
\includegraphics[width=0.35\textwidth]{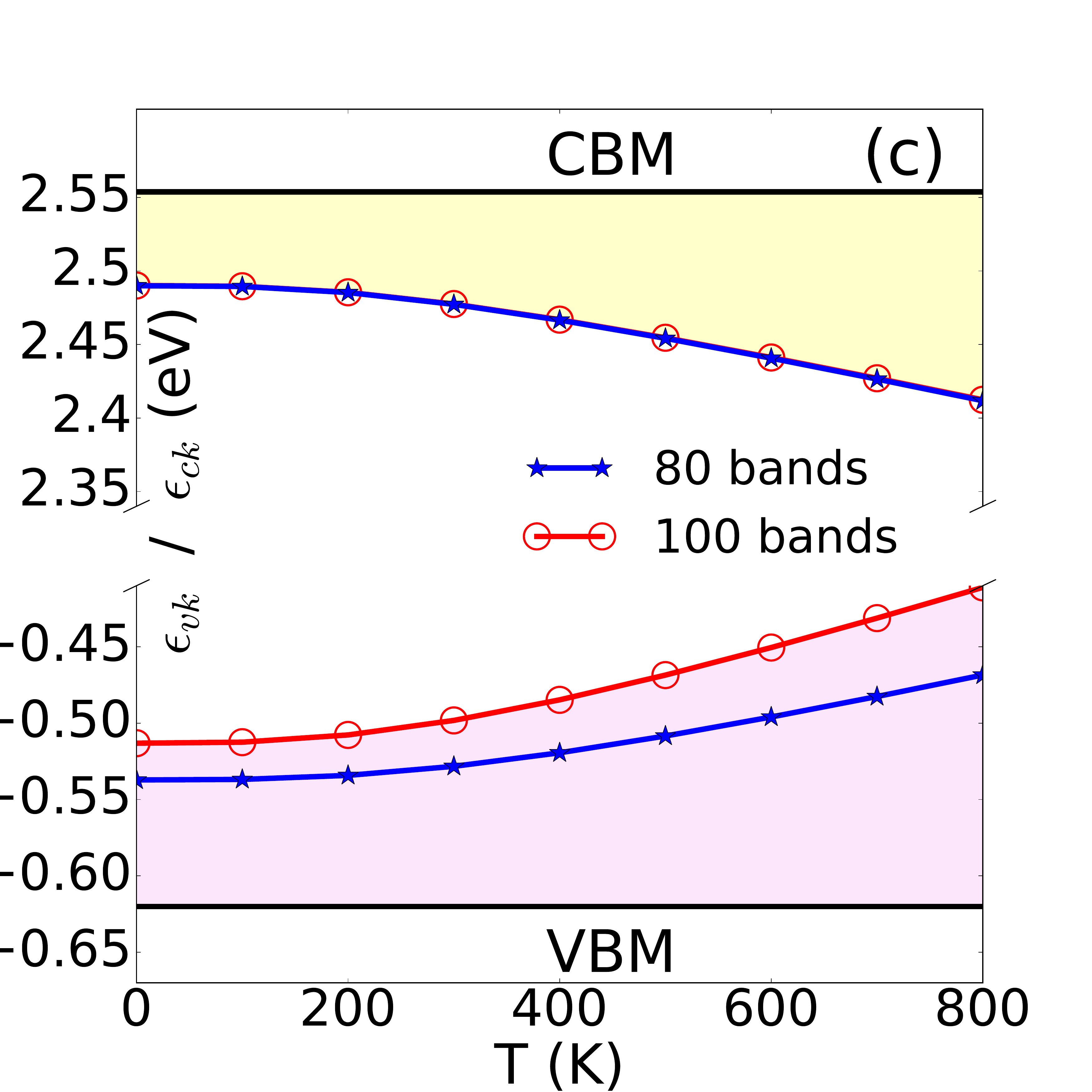}
\includegraphics[width=0.35\textwidth]{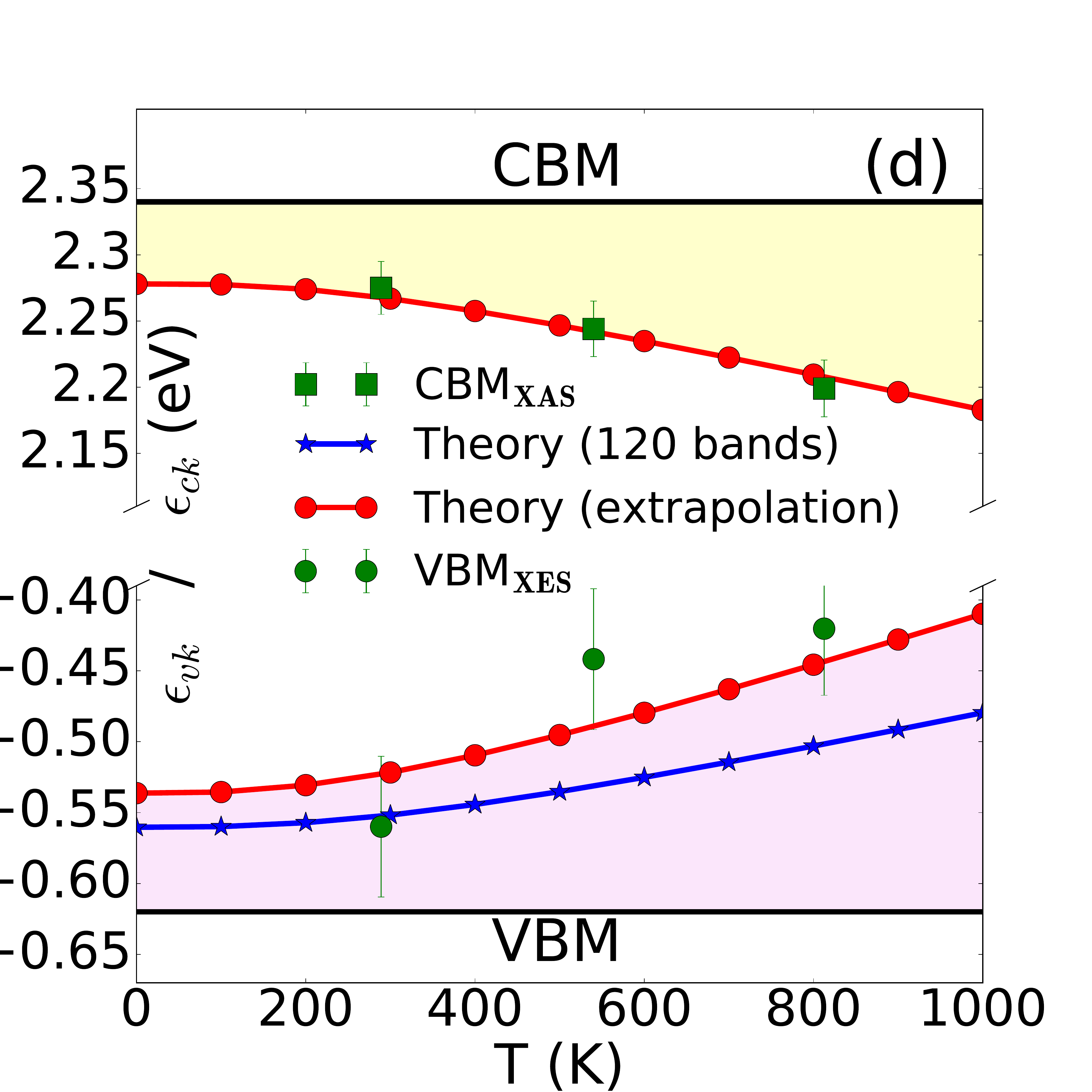}
\caption{\label{fig:conv} Calculated thermal evolution of indirect energy gaps as a function of the number of conduction bands in the calculation of electron-phonon coupling. (a) and (c) $4H$ SiC for the temperature range $0-800$~K, the effect of the growing number of unoccupied bands on the thermal evolution of band edges is shown. (b) and (d) $6H$ SiC for the temperature range $0-1000$~K, with explicit calculation of 120 bands and extrapolated values at 150 bands based on the $4H$ SiC data. 
In (a) and (b) the calculated curves are aligned to the experimental data at $T=0~\text{K}$ after having applied the GW correction to the indirect gap. 
The experimental data (green crosses) of the optical gaps are taken from Refs.~\onlinecite{Choyke1969, Pscajev2009}. The others (violet dots) have been extracted from RIXS spectra~(Ref.~\onlinecite{Miedema2014}), with correction of the core-hole exciton binding energy included. 
Thermal evolution of calculated VBM and CBM of (c) $4H$ SiC and (d) $6H$ SiC, where the latter is compared to the XAS and XES measurements (Ref.~\onlinecite{Miedema2014}) rescaled of 99.5~eV and 98.65~eV, respectively, in order to match the temperature evolution of CBM and VBM. The depedence of the thermal evolution of the individual bands as a function of the number of conduction bands in the calculation of electron-phonon coupling is analysed. Notice that on top of the electron-phonon correction the quasi-particle correction has been also added.
}
\end{figure}

We find that the calculated temperature shifts of the valence bands depends on the number of conduction bands used in the calculation of electron-phonon coupling whereas the temperature shifts of the conduction bands converge much faster with the number of conduction bands as shown in Fig.~\ref{fig:conv}(c). As can be inferred in Fig.~\ref{fig:conv}(a), the calculated temperature shifts in the valence band and the indirect band gap are absolutely convergent by using 100 bands in the calculation of the electron-phonon coupling for $4H$ SiC, as the 90 bands and 100 bands data are practically identical. This corresponds to about $6\times$ of the number of valence bands ($16$ bands without core electrons). As a consequence, the convergent number of bands should be around 150 for $6H$ SiC that has $24$ valence bands. However, it is technically prohibitive to carry out such calculations for $6H$ SiC. Thus, we carried out electron-phonon coupling calculations with non-converged 120 bands, which is only $5\times$ of the number of valence bands. We estimated the converged data by taking the difference in the calculated 80 bands data ($5\times$ of the number of valence bands in $4H$ SiC) and the convergent 100 bands data of $4H$ SiC. Then we extrapolated the convergent 150 bands numerical data by these differences at the given temperature in $6H$ SiC obtaining the thermal evolution of the indirect band gap and of VBM and CBM as shown in Fig.~\ref{fig:conv}(b)-(d).      

Next we turn to the calculated temperature shifts for the two lowest energy conduction bands around $M$ point. We found an energy separation of $89$~meV at DFT level between the two lowest conduction bands of $4H$-SiC, as reported in Table~I of the main manuscript. QP corrections shift down the second conduction band with respect to the first one of $3$~meV. The final inclusion of the electron-phonon interaction leads to a positive zero point energy correction of $2$~meV and $6$~meV at $800$~K. For $6H$ SiC the energy separation is $170$~meV at DFT level, it increases up to $180$~meV by the inclusion of QP effects, and then it gets to $183$~meV with zero point energy correction.   
%
%
\begin{figure}[h]
\includegraphics[width=0.45\textwidth]{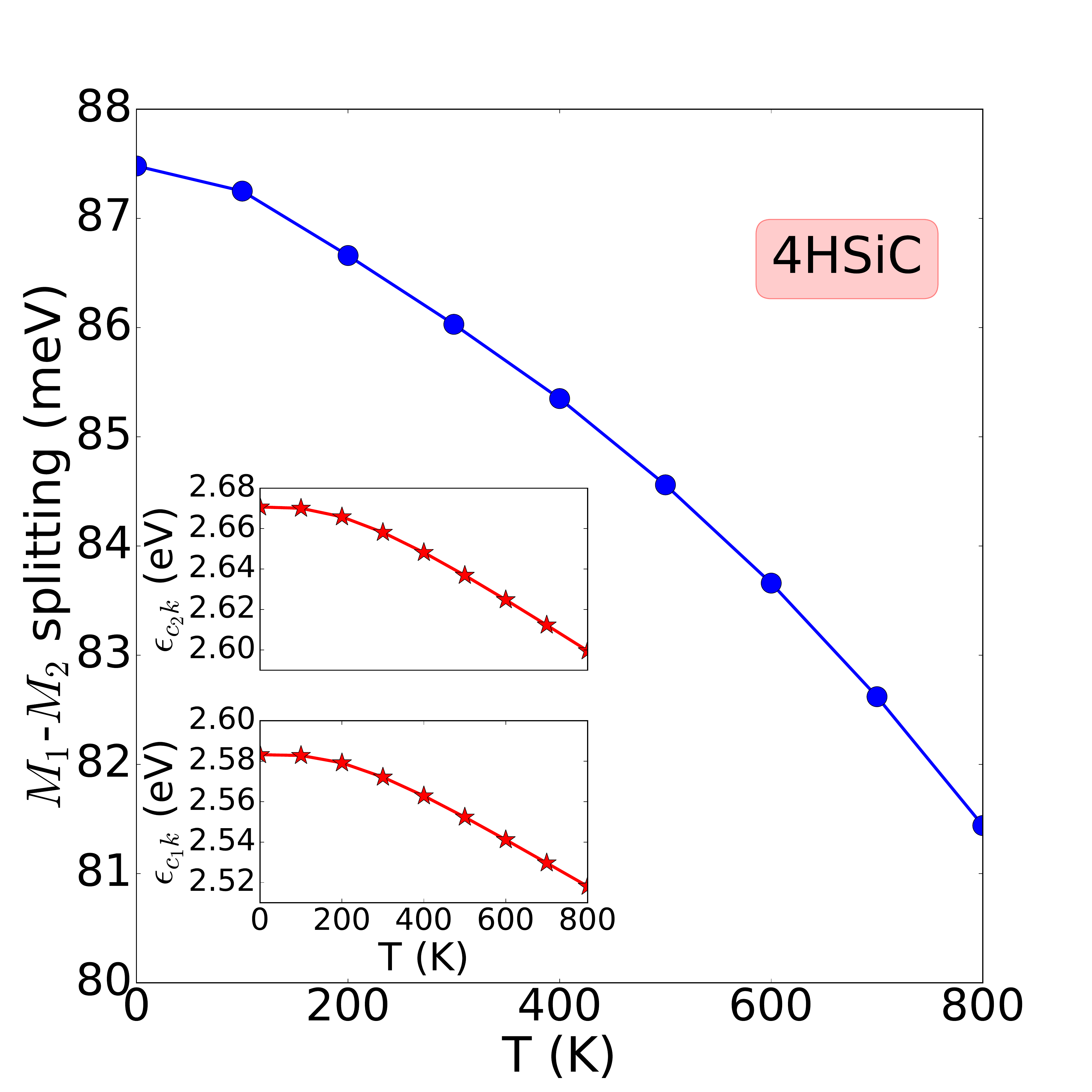}
\includegraphics[width=0.45\textwidth]{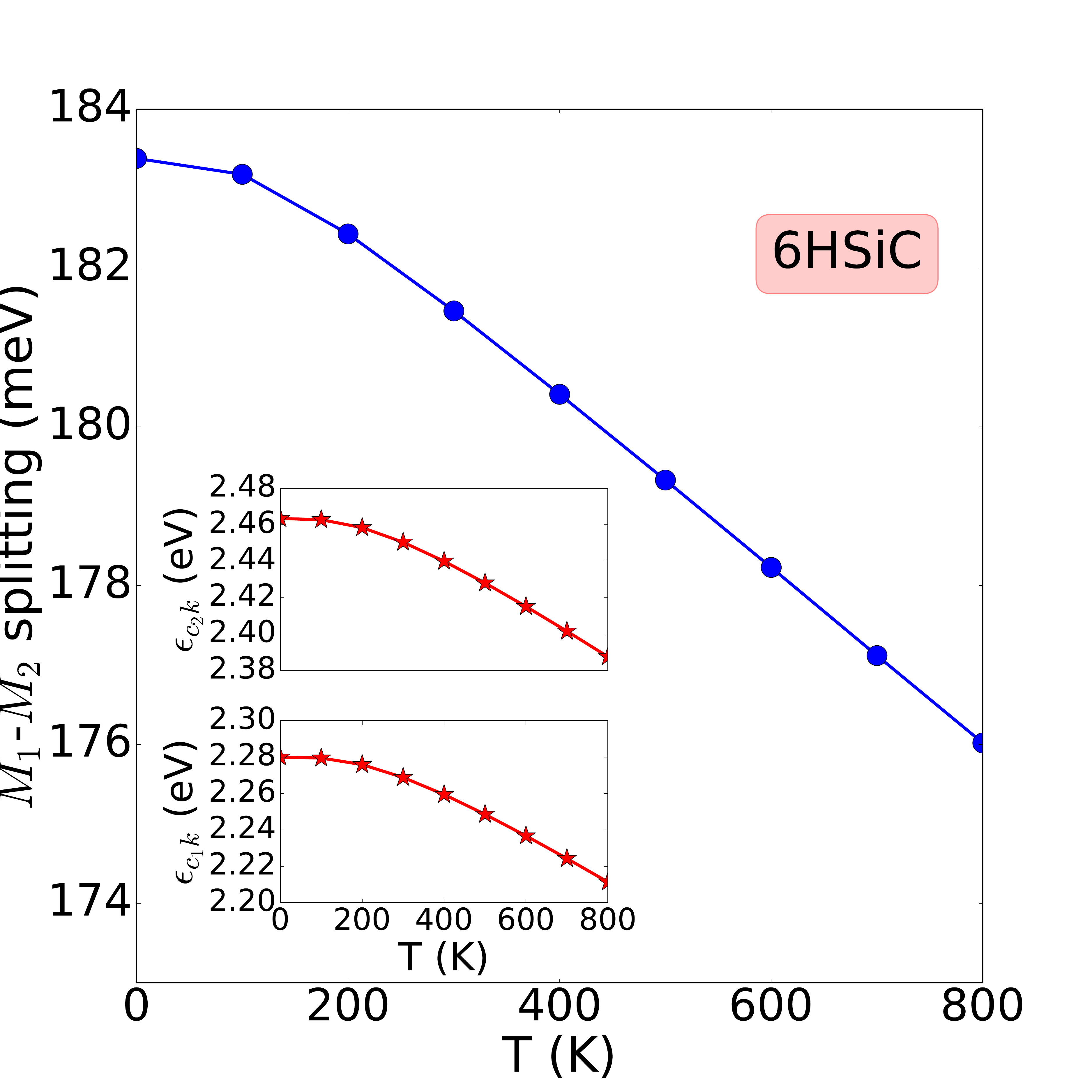}
\caption{\label{fig:Mbands}Temperature dependent energy gap between $M_1$ and $M_2$ conduction bands for $4H$ SiC and $6H$ SiC as obtained starting from DFT-LDA and QP corrected band structure. The experimental splitting at 2~K is about 144~meV in $4H$ SiC (Ref.~\onlinecite{Klahold2017}). } 
\end{figure}

%
%
%

\section{Additional data about conductivity of electrons in $4H$ SiC}
\begin{figure}
    \includegraphics[width=0.8\textwidth]{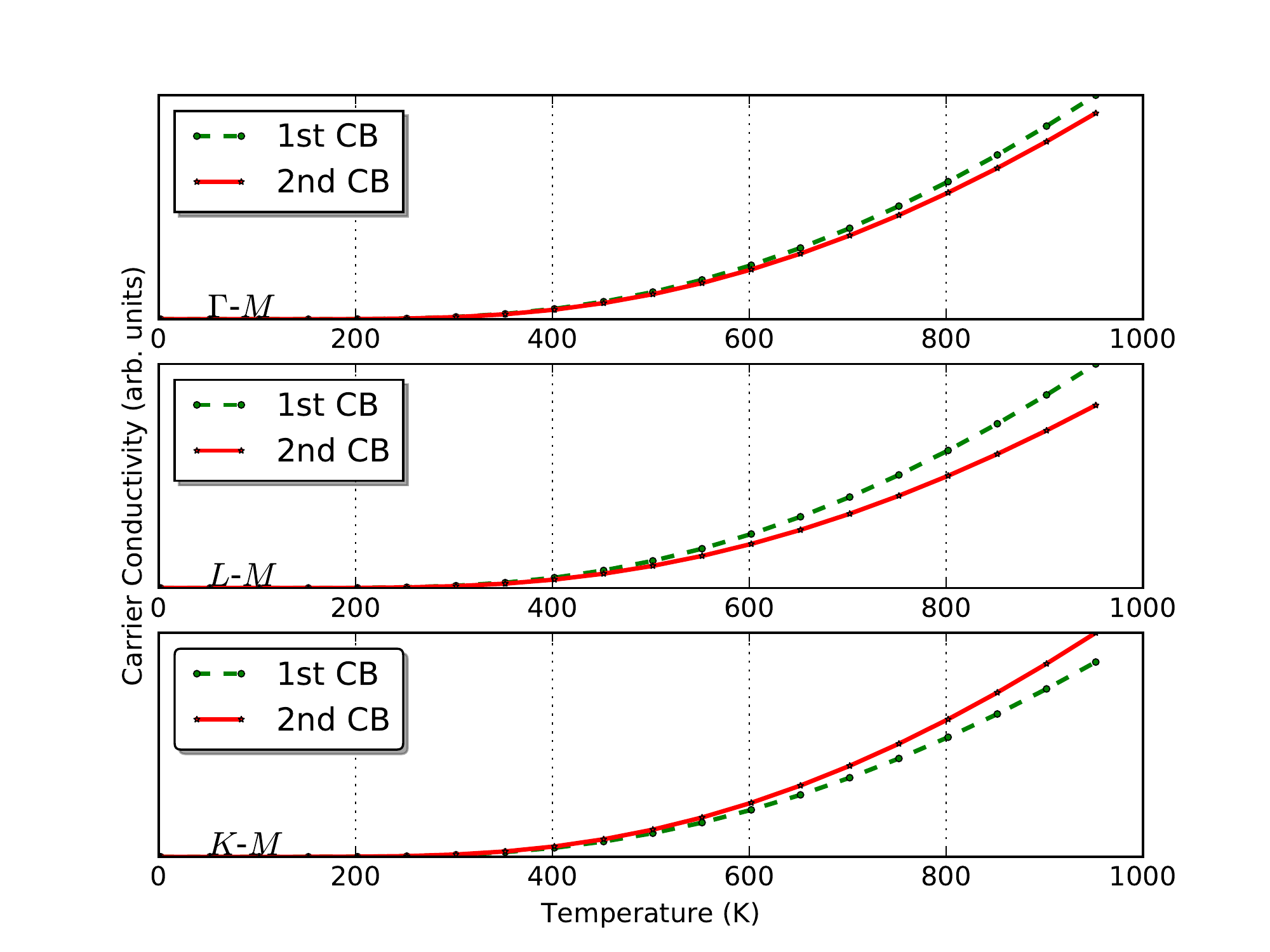}
    \caption{Carrier Conductivity for 120~meV donor level for the first and second conduction bands of $4H$ SiC and along the three directions $\Gamma$-M, L-M, K-M.}
    \label{fig:Conductivity_120meV}
\end{figure}

In extrinsic semiconductors, containing donor impurities, the presence of donor levels shifts the Fermi level from the middle of the energy gap toward the edge of the conduction band. Let us define the ionization temperature of the donor levels $T_d$ as
\begin{equation}
k_B T_d =\epsilon_d
\end{equation}
where $\epsilon_d$ is the binding energy of the donor levels, $k_B$ is the Boltzmann constant and the density $N_d$ of donor impurities is supposed to be uniform in the sample. Within the perspective of employing $4H$ SiC in devices operating at temperature of around 800 K, we expect that all the donor levels at $120$~meV below the bottom of the conduction band, are occupied and the chemical potential must be located in the energy range $E_d < \mu(T) < E_c$, where $E_d$ is the donor level and $E_c$ is the lowest conduction band position. In the low temperature regime, $T\ll T_d$ the temperature dependence of carrier conductivity is given by Refs.~\onlinecite{GrossoPastori, Fanfoni}
\begin{equation}
\sigma_{imp} \sim (\sqrt{N_c(T) \frac{N_d}{2}} e^{-\epsilon_d/2k_B T} ) \cdot e \cdot \frac{e}{m^*_c}  T^{3/2},
\end{equation}
being inversely proportional to the effective mass of the electron, which is the inverse of the second derivative of the conduction band along a given direction in the Brillouin-zone. The effective masses of the second conduction band ($M_2$) is generally greater (except along $K-M$ line) than of the first conduction band ($M_1$) (see Table~\ref{tab:EffMasses}). $N_d$ is the donor impurities density $\sim 10^{18}-10^{19} \frac{\# donors}{cm^{3}}$ and electronic charge $e=1.6\times 10^{-19} C$. 
The carrier conductivity as a function of temperature for binding energy donor levels equal to 120~meV in $4H$ SiC is represented in Fig.~\ref{fig:Conductivity_120meV}.

On the other hand for binding energy donor levels equal to 60~meV the carrier conductivity as a function of temperature is given by Refs.~\onlinecite{GrossoPastori, Fanfoni}
\begin{equation}
\sigma \sim N_d e \frac{e}{m^*_c} T^{3/2}.
\label{eq:conducibility_vs_T_sat_region}
\end{equation}

\begin{table}
\caption{\label{tab:EffMasses} $4H$ SiC effective masses of electrons in the conduction band minima in units of free electron mass $m_0$. Experimental data (exp.) are taken from Ref.~\onlinecite{Son1995}.}
\begin{ruledtabular}
\begin{tabular}[c]{ccccc}
Effective mass       &  direction        &    $M_1$       & $M_1$ (exp.) & $M_2$  \\ \hline
$m_{\bot}$   &  $m_{\Gamma M}$   &      0.48       &  0.42 &  0.67        \\
$m_{||}$ &  $m_{LM}$             &      0.27       &  0.29 &  0.61        \\ 
           &  $m_{KM}$           &      0.28       &       &  0.16        \\ 
\end{tabular}    
\end{ruledtabular}
\end{table}
 
\bibliographystyle{apsrev4-1}
\bibliography{references}